\documentclass[12pt]{article}

\usepackage[margin=1in]{geometry}

\def\real{\mathbb R}

\newcommand{\x}[1][j]{{X}^{(#1)}}
\newcommand{\z}[1][j]{{Z}^{(#1)}}
\newcommand{\lowerz}[1][j]{{z}^{(#1)}}
\newcommand{\hz}[1][j]{{\widehat{{Z}}}^{(#1)}}
\newcommand{\hblk}[1][j]{{\widehat{b}}^{(#1)}}
\newcommand{\degr}[1][j]{d^{(#1)}}
\newcommand{\del}[1][j]{\delta^{(#1)}}
\newcommand{\thet}[1][j]{{\theta}^{(#1)}}

\newcommand{\et}[1][j]{\eta^{(#1)}}
\newcommand{\het}[1][j]{{\widehat \eta}^{(#1)}}

\newcommand{\prob}[1][j]{{\pi}^{(#1)}}
\newcommand{\hprob}[1][j]{\widehat{\pi}^{(#1)}}

\renewcommand{\P}{\mathbb P}

\newcommand{\kk}[1][j]{{K}^{(#1)}}

\usepackage[figuresright]{rotating}
\usepackage{amsmath, amsfonts, amssymb}
\usepackage{algorithmic, algorithm, enumerate,color,hyperref, multirow, appendix, dsfont, xr, float, natbib}

\newtheorem{prop}{Proposition}

\title{Testing for Association in Multi-View Network Data}

\author{Lucy L. Gao$^{\dagger}$\footnote{Corresponding author: lucy.gao@uwaterloo.ca}~, Daniela Witten$^{\ddagger}$, and Jacob Bien$^{\circ}$ \\~\\
{\small $\dagger$ Department of Statistics and Actuarial Science, University of Waterloo} \\ 
{\small $\ddagger$ Departments of Statistics and Biostatistics, University of Washington} \\ 
{\small $\circ$ Department of Data Sciences and Operations, University of Southern California}  \\
}

\begin{document}
\maketitle

\begin{abstract}
In this paper, we consider data consisting of multiple networks, each comprised of a different edge set on a common set of nodes. Many models have been proposed for the analysis of such \emph{multi-view} network data under the assumption that the data views are closely related. In this paper, we provide tools for evaluating this assumption.  In particular, we ask: given two networks that each follow a stochastic block model, 
 is there an association between the latent community memberships of
 the nodes in the two networks? To answer this question, we extend the
 stochastic block model for a single network view to the two-view setting, and develop a new hypothesis test for the null hypothesis that the latent community memberships in the two data views are independent. We apply our test to
 protein-protein interaction data from the HINT database
 \citep{das2012hint}. We find evidence of a weak association between the latent community memberships of proteins defined with respect to binary interaction data
 and the latent community memberships of proteins defined with respect to co-complex association data. We also extend this proposal to the setting of a network with node covariates. The proposed methods extend readily to three or more network/multivariate data views. 
\end{abstract}

\section{Introduction}
A network consists of the pairwise relationships (edges) between objects of interest (nodes). For example, nodes could correspond to proteins, with edges representing physical interactions, or nodes could correspond to people, with edges representing social interactions. Of the many models for network data \citep{erdHos1960evolution, holland1981exponential, hoff2002latent}, one of the best known is the stochastic block model \citep{holland1983stochastic}, which assumes that nodes belong to latent communities, and that the probability of an edge between a pair of nodes is a function of their community memberships only.

\begin{figure}[h!]
(i) \fbox{\includegraphics[scale=0.35]{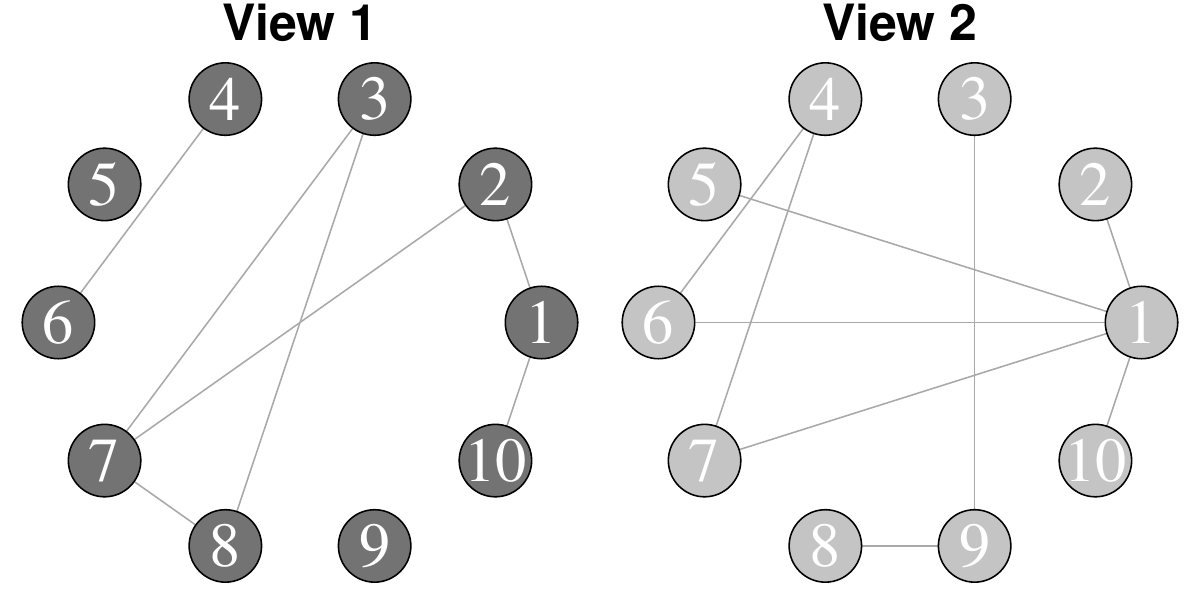}} \hspace{3mm}   (ii) \fbox{\includegraphics[scale=0.35]{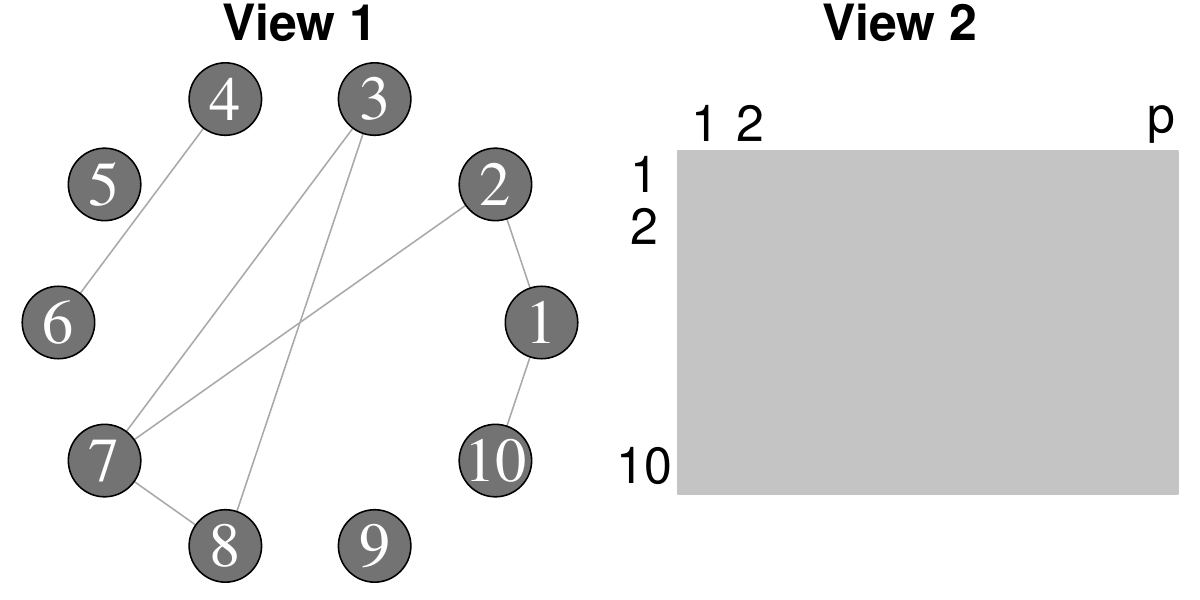}} 
\caption{\label{fig:toy} Two examples of multi-view data involving a network. (i) Two network views on $n = 10$ nodes. (ii) A network view and an $n \times p$ multivariate view on $n = 10$ nodes.} 
\end{figure}

Multiple sets of edges are often available on a common set of nodes, as is shown in Figure \ref{fig:toy}(i). Consider a pair of protein-protein interaction networks in which the nodes correspond to proteins. In one network, the edges represent physical interactions, and in the other, they represent co-membership in a protein complex. Another often-encountered scenario involves a single network, with a set of covariates corresponding to each node, as is shown in Figure \ref{fig:toy}(ii). For instance, we might have a social network along with $p$ demographic covariates for each member of the network. Both Figures \ref{fig:toy}(i) and \ref{fig:toy}(ii) are examples of the \emph{multi-view} data setting \citep{sun2013survey}. We will refer to the two networks in Figure \ref{fig:toy}(i), or the network and the covariates corresponding to the nodes in Figure \ref{fig:toy}(ii), as two \emph{data views}. 

Extensions of network models to the multi-view data setting
 \citep{fosdick2015testing, han2015consistent, gollini2016joint, binkiewicz2017covariate, salter2017latent, d2019latent} often
assume that the data views are closely related. For example, extensions of the stochastic block model typically assume that the latent communities within each network view are closely related \citep{han2015consistent, peixoto2015inferring, stanley2016clustering, binkiewicz2017covariate, stanley2019stochastic}. 

In this paper, we propose a test of the assumption that the latent communities are related. Why is this important? First of all, we should check whether two data views are in fact associated before we fit a model that relies on this assumption. Second, the relationship between the views may itself be of interest, and the test that we propose will allow us to assess this relationship. For example, such a tool can help shed light on whether the two distinct definitions of protein interactions capture similar versus complementary latent structures. Likewise, it can provide insight about whether peoples' social interactions and demographics are related. \citet{gao2020clusterings} investigated a similar problem for two multivariate data views, but did not consider the case where one or both views are networks.

To this end, we extend the stochastic block model to the multi-view network setting (Figure \ref{fig:toy}(i)) \emph{without} assuming that the network views are closely related. We then ask: are the latent communities within each network view associated? Similarly, for the case of a network view and a multivariate view (Figure \ref{fig:toy}(ii)), we model the network view with a stochastic block model and model the multivariate view with a finite mixture model, without assuming that the views are closely related. We then ask: are the latent communities within the network data view and the latent clusters within the multivariate data view associated? 

The rest of the paper is organized as follows. We review the stochastic block model in Section \ref{sec:sbm}. We extend the stochastic block model to two network data views in Section \ref{sec:networks-mod}, and develop a test for association between the latent communities within each view in Section \ref{sec:multi-sbm-test}. We develop a related test for the case of a network view and a multivariate view in Section \ref{sec:node}. We review related literature in Section \ref{sec:lit}, and explore the performance of our tests via simulation in Section \ref{sec:sim}. In Section \ref{sec:app}, we apply the test from Section \ref{sec:multi-sbm-test} to protein networks from the HINT database \citep{das2012hint}. Section \ref{sec:discuss} provides a discussion.

\section{The stochastic block model \citep{holland1983stochastic}}
In this section, we briefly review the stochastic block model (SBM) proposed by \citet{holland1983stochastic} for a single network; see \citet{matias2014modeling} for a detailed review. 
\label{sec:sbm}
\subsection{Model and notation} 
\label{sec:sbm-mod}
Let $X \in \{0, 1\}^{n \times n}$ be the adjacency matrix of an undirected, unweighted network with $n$ nodes and no self-loops, so that $X$ is symmetric and $X_{ii} = 0$ for $i = 1, 2, \ldots, n$. We assume that the nodes are partitioned into $K$ communities, with unobserved memberships given by a latent random vector $Z = (Z_1, \ldots, Z_n)$ with independent and identically distributed (i.i.d.) elements and $\P(Z_i = k) \equiv \pi_k$ for $\pi \in \Delta_+^K \equiv \left \{ \pi\in \real^{K} : 1_{K}^T\pi = 1, \pi_k > 0 \right \}$.
Conditional on $Z$, the edges are independently drawn from a Bernoulli distribution, with $\P[X_{ij} = 1 \mid Z] = \theta_{Z_i Z_j}$ for a symmetric matrix $\theta \in [0, 1]^{K \times K}$. It follows that
\begin{align} 
f(X \mid Z) = \prod \limits_{i=1}^n \prod \limits_{j=1}^{i-1} \left (\theta_{Z_i Z_j} \right )^{X_{ij}} \left ( 1 - \theta_{Z_i Z_j} \right )^{1 - X_{ij}}, \quad \mathbb{P}(Z = z) = \prod \limits_{i=1}^n \pi_{z_i}.\label{eq:SBM} 
\end{align} 

\subsection{Approximate pseudo-likelihood function} 
\label{sec:sbm-pseudo}
As a result of \eqref{eq:SBM}, the log-likelihood function for the SBM is given by 
\begin{align} 
\ell(\theta, \pi; X) \equiv \log \left ( \sum \limits_{z_1 = 1}^K \ldots \sum \limits_{z_n = 1}^K \left (\prod \limits_{i=1}^n \prod \limits_{j=1}^{i-1} \left (\theta_{z_i z_j} \right )^{X_{ij}} \left ( 1 - \theta_{z_i z_j} \right )^{1 - X_{ij}} \right ) \left ( \prod \limits_{i=1}^n \pi_{z_i} \right ) \right ). \label{eq:SBM-lik} 
\end{align} 
Equation \eqref{eq:SBM-lik} sums over $K^n$ terms, and is thus computationally intractable. Therefore,
\citet{amini2013pseudo} developed an approximate \emph{pseudo-likelihood} function, in the sense of \citet{besag1975statistical}. We briefly review this approach; see Appendix \ref{sec:approx} for a detailed review. 

Let $\widehat Z \in \{1, \ldots, K\}^n$ be the results of applying spectral clustering with perturbations \citep{amini2013pseudo} to $X$. Define $\widehat b \in \real^{n \times K}$ with rows $\widehat b_i$ and 
$\widehat b_{im} \equiv \sum \limits_{j=1}^n X_{ij} \mathds{1}\{\widehat{Z}_j = m\}$, 
and let $d = X 1_n$. Here, $\widehat b_{im}$ is the number of edges connecting the $i$th node to the $m$th estimated community in $\widehat Z$, and $d$ contains the degrees of the $n$ nodes. Let $\widehat R$ be the confusion matrix between $\widehat Z$ and $Z$, and define the $K \times K$ matrix $\eta = (\text{diag}(\theta \widehat{R} 1_K) )^{-1}\theta \widehat{R}$, with rows $\eta_1, \ldots, \eta_K \in \Delta_+^K$. 
Let $g(\cdot; N, q)$ denote the probability mass function of a Multinomial$(N, q_1, \ldots, q_K)$ random variable. \citet{amini2013pseudo} treated $\widehat Z$ and $\eta$ as fixed and showed that 
\begin{align} 
\widehat b \mid d, Z ~\dot \sim ~\prod \limits_{i=1}^n g \left (\widehat b_i ; d_i,  \eta_{Z_i} \right ), \label{eq:prodmulti} 
\end{align} 
where $\dot \sim$ denotes ``approximately distributed as". Ignoring any dependence between $Z$ and $d$, and marginalizing over $Z$ in \eqref{eq:prodmulti} 
to approximate the conditional distribution of $\widehat b$ given $d$, yields the following log-pseudo-likelihood function:
\begin{align} 
\label{eq:pseudo-log-lik}
\ell_{PL}(\eta, \pi; \widehat b \mid d) \equiv \sum \limits_{i=1}^n \log \left (  \sum \limits_{k=1}^K \pi_k g(\widehat b_i ; d_i, \eta_k) \right ).
\end{align} 
This can be viewed as the log-likelihood function of a finite mixture model (FMM; \citealt{mclachlan2004finite}) with $K$ components, of which the $k$th component has prior probability $\pi_k$ and density function $g(\widehat b_i; d_i, \eta_k)$.

\section{A stochastic block model for two network data views}
\label{sec:networks-mod}
In this section, we extend the SBM to the setting of two network data views, and derive approximate pseudo-likelihood functions for the proposed multi-view SBM.

\subsection{Model and notation} 
\label{sec:multi-sbm-mod}
Suppose that we have two network views on a common set of $n$ nodes, as in Figure \ref{fig:toy}(i), e.g. a binary network and a co-complex network on $n$ proteins. We assume that the networks are undirected, unweighted, and have no self-loops. Let $\x[1], \x[2] \in \{0, 1\}^{n \times n}$ be the symmetric adjacency matrices of the two networks, where $\x[l]_{ii} = 0$ for $i = 1, 2, \ldots, n$ and $l = 1, 2$.

We model $\x[1]$ with a SBM (Section \ref{sec:sbm-mod}) with $\kk[1]$ communities, and $\x[2]$ with a SBM with $\kk[2]$ communities. It follows from \eqref{eq:SBM} that for $l = 1, 2$, 
\begin{gather} 
f(\x[l] \mid \z[l]) = \prod \limits_{j=1}^n \prod \limits_{i=1}^{j-1} \left (\thet[l]_{\z[l]_i \z[l]_j} \right )^{\x[l]_{ij}} \left (1 -  \thet[l]_{\z[l]_i \z[l]_j} \right)  ^{1 - \x[l]_{ij}}, \quad 
\P(\z[l] = z^{(l)}) = \prod \limits_{i=1}^n \prob[l]_{z^{(l)}_i}, \label{eq:multi-sbm-latent} 
\end{gather}
for a symmetric matrix $\thet[l] \in [0, 1]^{\kk[l] \times \kk[l]}$ and $\prob[l]\in \Delta_+^{\kk[l]}$. Here, for $l = 1, 2$, $\z[l]$ represents the latent community memberships for the $n$ nodes within the $l$th network data view. We assume that the $n$ pairs $\{(\z[1]_i, \z[2]_i)\}_{i=1}^n$ are i.i.d. and that $\x[1] \perp \x[2] \mid \z[1], \z[2]$.

The following result allows us to parameterize the joint distribution of $\z[1]$ and $\z[2]$.
\begin{prop}[\citealt{gao2020clusterings}]  
\label{prop:reparam}
Consider two categorical random variables $A$ and $B$ with $K$ and $K'$ levels, respectively, and with $\P(A = k) = \pi_k$ and $\P(B = k') = \pi'_{k'}$, for $\pi \in \Delta_+^{K}$ and $\pi' \in \Delta_+^{K'}$. Then, there exists a unique matrix $C \in \mathcal{C}_{\pi, \pi'}$ such that 
$\P(A = k, B= k') =  \pi_k \pi'_{k'} C_{kk'},$
where $\mathcal{C}_{\pi, \pi'} \equiv \{C \in \real^{K \times K'}: ~ C_{kk'} \geq 0,~  C  \pi' = 1_{K}, ~C^T \pi = 1_{K'}  \}. $
\end{prop}

It follows from applying Proposition \ref{prop:reparam} to each of the $n$ pairs of categorical variables $\{(\z[1]_i, \z[2]_i)\}_{i=1}^n$ that there exists a unique $\kk[1] \times \kk[2]$ matrix $C \in \mathcal{C}_{\prob[1], \prob[2]}$ such that
\begin{align} 
\P(\z[1]=\lowerz[1], \z[2]=\lowerz[2])&= \prod \limits_{i=1}^n \P(\z[1]_i = \lowerz[1]_i, \z[2]_i = \lowerz[2]_i) = \prod \limits_{i=1}^n \prob[1]_{\lowerz[1]_i} \prob[2]_{\lowerz[2]_i} C_{\lowerz[1]_i\lowerz[2]_i},
 \label{eq:mv-subgroup-z}
\end{align} 
where the first equality follows from the independence of the $n$ pairs $\{(\z[1]_i, \z[2]_i)\}_{i=1}^n$. Here, $C_{kk'} = \frac{\P(\z[1]_i = k, \z[2]_i = k')}{\P(\z[1]_i = k)\P(\z[2]_i = k')}$ describes the dependence between the $k$th community in the first view and the $k'$th community in the second view, with $C_{kk'} = 1$ indicating independence, $C_{kk'} < 1$ indicating negative dependence, and $C_{kk'} > 1$ indicating positive dependence. 

\subsection{Approximate pseudo-likelihood function} 
\label{sec:multi-sbm-pseudo} 
The log-likelihood function of  model \eqref{eq:multi-sbm-latent}--\eqref{eq:mv-subgroup-z} is given by 
\begin{gather} 
\ell(\thet[1], \thet[2], \prob[1], \prob[2], C; \x[1], \x[2])\label{eq:SBM-multi-lik} \\ 
\equiv \log \left ( \scriptstyle  \sum \limits_{\lowerz[1]_1 = 1}^{\kk[1]} \ldots \sum \limits_{\lowerz[1]_n = 1}^{\kk[1]}  \sum \limits_{\lowerz[2]_1 = 1}^{\kk[2]} \ldots \sum \limits_{\lowerz[2]_n = 1}^{\kk[2]}  \left (  \prod \limits_{l=1}^2 \prod \limits_{i=1}^n \prod \limits_{j=1}^{i-1} \left (\thet[l]_{\lowerz[l]_i \lowerz[l]_j} \right )^{\x[l]_{ij}} \left ( 1 - \thet[l]_{\lowerz[l]_i \lowerz[l]_j} \right )^{1 - \x[l]_{ij}} \right ) \left (  \prod \limits_{i=1}^n \prob[1]_{\lowerz[1]_i}\prob[2]_{\lowerz[2]_i} C_{\lowerz[1]_i \lowerz[2]_i} \right ) \right ).  \nonumber 
\end{gather} 
Equation \eqref{eq:SBM-multi-lik} is computationally intractable, because it involves summing over $(\kk[1] \kk[2])^n$ terms. Thus, we will derive an approximate pseudo-likelihood function for model \eqref{eq:multi-sbm-latent}--\eqref{eq:mv-subgroup-z}. For $l = 1,2$, let $\hz[l] \in \{1, \ldots, \kk[l]\}^n$ be the results of applying spectral clustering with perturbations \citep{amini2013pseudo} to $\x[l]$, let $\hblk[l]$ be the $n \times \kk[l]$ matrix defined by $\hblk[l]_{im} = \sum \limits_{i=1}^n \x[l]_{ij} \mathds{1} \{\hz[l]_j = m\}$ 
and let $\degr[l] = \x[l] 1_n$. Here, for the $l$th network, $\hblk[l]_{im}$ is the number of edges connecting the $i$th node to the $m$th estimated community, and $\degr[l]_i$ is the degree of the $i$th node. 
We write

\begin{gather} 
f({\scriptstyle \hblk[1], \hblk[2] } \mid {\scriptstyle  \degr[1], \degr[2], \z[1], \z[2]})  = \frac{f({\scriptstyle \hblk[1], \hblk[2], \degr[1], \degr[2]} \mid {\scriptstyle \z[1], \z[2]})}{f({\scriptstyle \degr[1], \degr[2] }\mid {\scriptstyle \z[1], \z[2]})}  = \textstyle \prod \limits_{l=1}^2 \frac{f(\hblk[l], \degr[l] \mid \z[l])}{f(\degr[l] \mid \z[l])}  = \textstyle \prod \limits_{l=1}^2 f(\hblk[l] \mid \degr[l], \z[l]), \label{eq:approx1}
\end{gather} 
where the first and third equalities follow from the definition of a conditional density, and the second equality follows from the fact that $\x[1] \perp \x[2] \mid \z[1], \z[2]$ and $\x[1] \perp \z[2] \mid \z[1]$ and $\x[2] \perp \z[1] \mid \z[2]$ (Section \ref{sec:multi-sbm-mod}). Let $\widehat R^{(l)}$ be the confusion matrix between $\hz[l]$ and $\z[l]$ and let $\et[l] = (\text{diag}(\thet[l] \widehat{R}^{(l)} 1_{\kk[l]}) )^{-1}\thet[l] \widehat{R}^{(l)}$. As in \citet{amini2013pseudo}, we treat $\hz[l]$ and $\et[l]$ as fixed, and apply \eqref{eq:prodmulti} in Section \ref{sec:sbm-pseudo} to approximate $f(\hblk[l] \mid \z[l], \degr[l])$ 
in \eqref{eq:approx1}, which yields
\begin{align} 
f(\hblk[1], \hblk[2] \mid \degr[1], \degr[2], \z[1], \z[2]) \approx \prod \limits_{l=1}^2 \prod \limits_{i=1}^n g(\hblk[l]_i; \degr[l]_i, \et[l]_{\z[l]_i}). \label{eq:approx2}
\end{align} 
Ignoring any dependence between $(\degr[1], \degr[2])$ and $(\z[1], \z[2])$ and marginalizing over the latent community memberships $\z[1]$ and $\z[2]$ in \eqref{eq:approx2} to approximate the conditional distribution of $\hblk[1]$ and $\hblk[2]$ given $\degr[1]$ and $\degr[2]$ yields the following log-pseudo-likelihood function: 
\begin{gather} 
\ell_{PL} (\et[1], \et[2], \prob[1], \prob[2], C; ~\hblk[1], \hblk[2] \mid \degr[1], \degr[2])\nonumber \\ 
\equiv \textstyle \sum \limits_{i=1}^n \log \left ( \sum \limits_{k=1}^{\kk[1]} \sum \limits_{k'=1}^{\kk[2]} \prob[1]_k\prob[2]_{k'} C_{kk'} g(\hblk[1]_i; \degr[1]_i, \et[1]_k) g(\hblk[2]_i; \degr[2]_i, \et[2]_{k'}) \right ). \label{eq:pseudo-log-lik-joint}
\end{gather} 
This closely resembles the log-likelihood function of the finite mixture model for two multivariate data views from \citet{gao2020clusterings}.

\section{Are two network views' community memberships associated?} 
\label{sec:multi-sbm-test}
Recall from \eqref{eq:mv-subgroup-z} that $\P(\z[1] = \lowerz[1], \z[2] = \lowerz[2]) = \prod \limits_{i=1}^n \prob[1]_{\lowerz[1]_i} \prob[2]_{\lowerz[2]_i} C_{\lowerz[1]_i\lowerz[2]_i}$, where $C \in \mathcal{C}_{\prob[1], \prob[2]}$, defined in Proposition \ref{prop:reparam}. It follows from the definition of $\P(\z[l] = \lowerz[l])$ in \eqref{eq:multi-sbm-latent} that 
\begin{align*} 
\P(\z[1] = \lowerz[1], \z[2] = \lowerz[2]) = \P(\z[1] = \lowerz[1]) \P(\z[2] = \lowerz[2])
\end{align*} 
if and only if $C = 1_{\kk[1]} 1_{\kk[2]}^T$. Thus, testing the null hypothesis of independence between the latent community memberships $\z[1]$ and $\z[2]$ amounts to testing $H_0: C = 1_{\kk[1]} 1_{\kk[2]}^T$.  

\subsection{The $P^2$LRT statistic} 
\label{sec:test-stat}
To test $H_0: C = 1_{\kk[1]} 1_{\kk[2]}^T$, one might consider using a likelihood ratio test. The likelihood ratio test statistic is of the form 
\begin{align*} 
\underset{\thet[1], \thet[2], \prob[1], \prob[2], C}{\max} {\scriptstyle \ell(   \et[1], \et[2], \prob[1], \prob[2], C;  \x[1], \x[2]) } - \max_{\et[1], \et[2], \prob[1], \prob[2]} \scriptstyle { \ell(   \thet[1], \thet[2], \prob[1], \prob[2], 1_{\kk[1]} 1_{\kk[2]}^T; \x[1], \x[2])}, 
\end{align*} 
where the log-likelihood function $\ell$ is defined in \eqref{eq:SBM-multi-lik}. Unfortunately, recall from Section \ref{sec:multi-sbm-pseudo} that \eqref{eq:SBM-multi-lik} is computationally intractable because it involves summing over $(\kk[1] \kk[2])^n$ terms. We could replace the log-likelihood functions $\ell$ with log-pseudo-likelihood functions $\ell_{PL}$, defined in \eqref{eq:pseudo-log-lik-joint}. This leads to a test statistic of the form 
\begin{align} 
\log \Lambda &\equiv \max_{\et[1], \et[2], \prob[1], \prob[2], C} ~{\textstyle \ell_{PL}(   \et[1], \et[2], \prob[1], \prob[2], C; ~\hblk[1], \hblk[2] \mid \degr[1], \degr[2]}) ~ - \nonumber \\ 
&~~~\max_{\et[1], \et[2], \prob[1], \prob[2]}~ {\textstyle \ell_{PL}(   \et[1], \et[2], \prob[1], \prob[2], 1_{\kk[1]} 1_{\kk[2]}^T; ~\hblk[1], \hblk[2] \mid \degr[1], \degr[2]})  \label{eq:plrt}. 
\end{align} 
However, $\ell_{PL}$ is a non-concave function of its arguments, and so no algorithms are available to exactly compute the two terms in \eqref{eq:plrt} --- they can at best be approximated via local maxima. Taking  the difference between two local maxima can lead to undesirable behavior;
 for example, $\log \Lambda$ can be negative.

To overcome this problem, we take a different approach, motivated by the fact that each data view $\x[l]$ marginally follows a SBM 
(Section \ref{sec:multi-sbm-mod}). Rather than estimating  the parameters $\et[1], \et[2], \prob[1], \prob[2]$ and $C$ by maximizing the log-pseudo-likelihood function for the multi-view SBM \eqref{eq:pseudo-log-lik-joint}, we first estimate $\et[1],\prob[1]$ and $\et[2],\prob[2]$ by maximizing the log-pseudo-likelihood function for the SBM \eqref{eq:pseudo-log-lik} for each view separately. Since \eqref{eq:pseudo-log-lik} can be viewed as the log-likelihood function of a FMM (Section \ref{sec:multi-sbm-pseudo}), it can be maximized using the expectation-maximization (EM; \citealt{dempster1977maximum}) algorithm for fitting FMMs \citep{mclachlan2007algorithm}. 
We then plug these estimates into \eqref{eq:plrt}, yielding the test statistic
\begin{align}
\log \widetilde \Lambda  &\equiv  \max_{C \in \mathcal{C}_{\hprob[1], \hprob[2]}}\ell_{PL}(   \het[1], \het[2], \hprob[1], \hprob[2], C; ~\hblk[1], \hblk[2] \mid \degr[1], \degr[2]) - \nonumber \\ 
&~~\ell_{PL}( \het[1], \het[2], \hprob[1], \hprob[2], 1_{\kk[1]} 1_{\kk[2]}^T; ~\hblk[1], \hblk[2] \mid \degr[1], \degr[2]). \label{eq:genp2lrt}  
\end{align} 
Computing \eqref{eq:genp2lrt} requires maximizing the first term with respect to $C$, i.e. to compute
\begin{align} 
\widehat C \equiv \underset{C \in \mathcal{C}_{\hprob[1], \hprob[2]}}{\arg \max} ~ \ell_{PL}(   \het[1], \het[2], \hprob[1], \hprob[2], C; ~\hblk[1], \hblk[2] \mid \degr[1], \degr[2]), \label{eq:pseudoCopt}
\end{align} 
where $\mathcal{C}_{\cdot, \cdot}$ is defined in Proposition \ref{prop:reparam}. Because the objective of \eqref{eq:pseudoCopt} is a concave function of $C$,  $\widehat{C}$ can be obtained using techniques from convex optimization. (In particular, we use an exponentiated gradient descent algorithm \citep{kivinen1997exponentiated} developed in \citet{gao2020clusterings} for maximizing concave functions of $C$ under the constraint that $C \in \mathcal{C}_{\hprob[1], \hprob[2]}$; the complexity of each iteration is $\mathcal{O}(nK^{(1)}K^{(2)})$. )
 This means that \eqref{eq:genp2lrt} completely overcomes the challenges associated with the test statistic \eqref{eq:plrt}; for example, \eqref{eq:genp2lrt} cannot be negative. Furthermore, results from \citet{liang1996asymptotic} and \citet{chen2010asymptotic} suggest that performing a partial maximization over the parameters (as in \eqref{eq:genp2lrt}) rather than a full maximization (as in \eqref{eq:plrt}) does not lead to an appreciable loss in power when $n$ is large. 

 We refer to $\log \widetilde \Lambda$ in \eqref{eq:genp2lrt} as a {\em pseudo-pseudo-likelihood ratio test} ($P^2$LRT) statistic. In the name $P^2$LRT, the term ``pseudo'' is used in two different senses: the first is because we use the pseudo-likelihood function $\ell_{PL}$ in place of the likelihood function, and the second is because we do not perform a full joint maximization over $(\et[1], \et[2], \prob[1], \prob[2], C) $. 

We summarize the procedure for computing the $P^2$LRT statistic in Algorithm \ref{alg:est}.

\begin{algorithm}[h!]
\caption{\label{alg:est} Computing the $P^2$LRT statistic $\log \widetilde \Lambda$ defined in \eqref{eq:genp2lrt}}
\begin{enumerate}[1.]
\item For $l = 1, 2$:
\begin{enumerate}[i.] 
\item Compute $\degr[l] = \x[l] 1_n$. Apply spectral clustering with perturbations \citep{amini2013pseudo} to $\x[l]$ to obtain $\hz[l]$, and compute $\hblk[l]$ according to $\hblk[l]_{im} = \sum \limits_{i=1}^n \x[l]_{ij} \mathds{1} \{\hz[l]_j = m\}$. 
\item Maximize $\ell_{PL}(\et[l], \prob[l]; \hblk[l] \mid \degr[l])$, where $\ell_{PL}$ is defined in \eqref{eq:pseudo-log-lik}, and denote the maximizers by $\het[l]$ and $\hprob[l]$. This can be done using the EM algorithm for fitting FMMs \citep{mclachlan2007algorithm}. 
\end{enumerate}
\item Compute $\widehat C$ according to \eqref{eq:pseudoCopt}: 
\begin{enumerate}[i.]
\item Define matrices $\widehat g^{(1)}\in\mathbb R^{n\times
K^{(1)}}$ and  $\widehat g^{(2)}\in\mathbb R^{n\times
K^{(2)}}$ with elements
$  \widehat g^{(1)}_{ik}=  g\left(\hblk[1]_i ; \degr[1]_i, \widehat \eta^{(1)}_k \right)$ and $\widehat g^{(2)}_{ik'}= g\left(\hblk[2]_i ; \degr[2]_i, \widehat \eta^{(2)}_{k'} \right).$
\item Fix a step size $s>0$, and let $\widehat{C}^1 = 1_{K^{(1)}} 1_{K^{(2)}}^T$.  For $t=1,2,\ldots$ until
    convergence: 
    \begin{enumerate}[a.] 
    \item Define $O_{kk'} =   \widehat{C}_{kk'}^t  \exp\{sG_{kk'} - 1\},$ where $G_{kk'}=\sum \limits_{i=1}^n\frac{\widehat g^{(1)}_{ik}\widehat g^{(2)}_{ik'}}{[\widehat g^{(1)}_i]^T \mathrm{diag}(\widehat \pi^{(1)}) \widehat{C}^t \mathrm{diag} (\widehat \pi^{(2)}) \widehat g^{(2)}_i}.$
\item  Let $u^0 = 1_{K^{(2)}}$ and $v^0 = 1_{K^{(1)}}$. For $t' = 1, 2, \ldots$, until convergence:
$$u^{t'} = \frac{1_{K^{(2)}}}{O^T \mathrm{diag}(\widehat \pi^{(1)}) v^{t' - 1}}, \quad v^{t'} = \frac{1_{K^{(1)}}}{O \mathrm{diag}(\widehat \pi^{(2)}) u^{t'}},$$ 
where the fractions denote element-wise vector division.
\item Let $u$ and $v$ be the vectors to which $u^{t'}$ and $v^{t'}$ converge. Let $\widehat{C}^{t+1}_{kk'}=  u_k O_{kk'} v_{k'}. $
  \end{enumerate}
  \item Let $\widehat{C}$ denote the matrix to which $\widehat{C}^t$ converges.
  \end{enumerate}
  \item Compute $\log \widetilde \Lambda$ according to \eqref{eq:genp2lrt},  where $\ell_{PL}$ is defined in \eqref{eq:pseudo-log-lik-joint}.
\end{enumerate} 
\end{algorithm} 

\subsection{Approximating the null distribution} 
\label{sec:null-approx}
Under the  null hypothesis that the community memberships $\z[1]$ and $\z[2]$ are independent, i.e. under $H_0: C = 1_{\kk[1]} 1_{\kk[2]}^T$, we can write the joint density of $\x[1]$ and $\x[2]$ as
\begin{align*} 
f(\x[1], \x[2]) &= \mathbb{E}_{\z[1], \z[2]} [ f(\x[1], \x[2] \mid \z[1], \z[2]) ] \\ 
&= \mathbb{E}_{\z[1], \z[2]} [ f(\x[1] \mid \z[1])f(\x[2] \mid \z[2]) ]  \\ 
&= \mathbb{E}_{\z[1]} [ f(\x[1] \mid \z[1])] \mathbb{E}_{\z[2]} [ f(\x[2] \mid \z[2]) ]= f(\x[1])f(\x[2]), 
\end{align*} 
where the second equality follows from the fact that $\x[1] \perp \x[2] \mid \z[1], \z[2]$ and $\x[1] \perp \z[2] \mid \z[1]$ and $\x[2] \perp \z[2] \mid \z[1]$ (Section \ref{sec:multi-sbm-mod}). Thus, under $H_0: C = 1_{\kk[1]} 1_{\kk[2]}^T$, the joint distribution of  $\x[1]$ and $\x[2]$ is invariant under permutation of the node labels $\{1, 2, \ldots, n\}$ in either network. It follows that we can approximate the null distribution of the $P^2$LRT statistic $\log \widetilde \Lambda$ defined in \eqref{eq:genp2lrt} by taking $M$ random permutations of the node labels in the second network, and comparing the observed value of $\log \widetilde \Lambda$ to its empirical distribution in the permuted data. Since $\het[1], \het[2]$, $\hprob[1]$, and $\hprob[2]$ are invariant to permutation, we only need to compute $\widehat C$ for each permutation. This is another advantage of the $P^2$LRT statistic $\log \widetilde \Lambda$ in \eqref{eq:genp2lrt} over $\log \Lambda$ in \eqref{eq:plrt}: if we had used $\log \Lambda$, then we would need to estimate $\et[1], \et[2], \prob[1], \prob[2],$ and $C$ for each permutation. Details of the testing procedure are in Algorithm \ref{alg:test}. In Step 3 of Algorithm \ref{alg:test}, we add 1 to the numerator and the denominator of the permutation p-value to ensure that the p-value is never exactly zero \citep{belinda2010permutation}. 

\begin{algorithm}[h!]
\caption{\label{alg:test} $P^2$LRT for testing $H_0: C = 1_{\kk[1]} 1_{\kk[2]}^T$} 
\begin{enumerate}[1.] 
\item Apply Algorithm \ref{alg:est} to compute $\hblk[1], \hblk[2], \degr[1], \degr[2]$, and the $P^2$LRT statistic $\log \widetilde \Lambda$ in \eqref{eq:genp2lrt}. 
 \item  For $m=1,\ldots, M$,  where $M$ is the number of random permutations:
  \begin{enumerate}[i.]
  \item Apply the same permutation to the rows of $\hblk[2]$ and the elements of $\degr[2]$ to compute $\hblk[2, *m]$ and $\degr[2, *m]$.
  \item Replace $\hblk[2], \degr[2]$  with $\hblk[2, *m], \degr[2*, m]$ in Step 2 of Algorithm \ref{alg:est} to compute $\widehat C^{(*m)}$. 
  \item Replace $\hblk[2]$, $\degr[2]$, and $\widehat C$ with $\hblk[2, *m]$, $\degr[2, *m]$, and $\widehat C^{(*m)}$ in  \eqref{eq:genp2lrt} to compute $ \log \widetilde \Lambda^{(*m)}$.
  \end{enumerate}
  \item The p-value for testing $H_0: C = 1_{\kk[1]} 1_{\kk[2]}^T$ is given by $\frac{\sum \limits_{m=1}^M \mathds{1}  \left\{ \log \Lambda \leq  \log \Lambda^{(*m)}  \right\}  + 1}{M+1}.$
  \end{enumerate}
\end{algorithm} 

When we reject $H_0: C = 1_{\kk[1]} 1_{\kk[2]}^T$, it is often of interest to investigate the strength and location of the dependence between views. Recall from Section \ref{sec:multi-sbm-mod} that $C_{kk'}$ measures the dependence between the $k$th community in the first view and the $k'$th community in the second  view. Thus, we can gain insight into the strength and location of the dependence between the communities in the two data views by examining $\widehat C_{kk'}$ defined in \eqref{eq:pseudoCopt}.

\section{Extension to a network view and a multivariate view} 
\label{sec:node}
In this section, we develop a test of association between latent communities in a network view and latent clusters in a multivariate view.

\subsection{Model and notation}
\label{sec:node-mod}
We now propose an extension of the SBM to an undirected network view, $X \in \{0, 1\}^{n \times n}$, and a multivariate view, $Y \in \real^{n \times p}$. We assume that the network is undirected with no self-loops, so that $X$ is symmetric and $X_{ii} = 0$ for $i = 1, 2, \ldots, n$.  We model $X$ with a SBM (Section \ref{sec:sbm-pseudo}) with $\kk[1]$ communities and we model the rows of $Y$ with a finite mixture model \citep{mclachlan2004finite} with $\kk[2]$ clusters, so that 
\begin{align} 
f(X \mid \z[1]) = \prod \limits_{j=1}^n \prod \limits_{i=1}^{j-1} (\theta_{\z[1]_i \z[1]_j})^{X_{ij}} (1 -  \theta_{\z[1]_i \z[1]_j})^{1 - X_{ij}}, \quad f(Y \mid \z[2]) = \prod \limits_{i=1}^n \phi(Y_i; \gamma_{\z[2]_i}),  \label{eq:mv-node-latent}
\end{align} 
where $\phi(\cdot; \gamma)$ is a density parameterized by $\gamma$, and for $l = 1, 2$, the latent random vector $\z[l] = (\z[l]_1, \ldots, \z[l]_n)$ has i.i.d. elements with $\P(\z[l]_i = k) = \prob[l]_k$ for $\prob[l] \in \Delta^{\kk[l]}_+$. Here, $\z[1]$ represents the latent community memberships in the network view, and $\z[2]$ represents the latent cluster memberships in the multivariate view. We assume that the $n$ pairs $\{(\z[1]_i, \z[2]_i)\}_{i=1}^n$ are i.i.d., and that $X \perp Y \mid \z[1], \z[2]$. Thus, as in Section \ref{sec:multi-sbm-mod}, it follows from Proposition \ref{prop:reparam} that there exists $C \in \mathcal{C}_{\prob[1], \prob[2]}$ such that 
\begin{align} 
\P(\z[1]=\lowerz[1], \z[2]=\lowerz[2])= \prod \limits_{i=1}^n \prob[1]_{\lowerz[1]_i} \prob[2]_{\lowerz[2]_i} C_{\lowerz[1]\lowerz[2]}, \label{eq:mv-node-subgroup-z}
\end{align} 
where $C_{kk'}$ describes the dependence between the $k$th community in the network view and the $k'$th cluster in the multivariate view.

\subsection{Approximate pseudo-likelihood function} 
The multi-view log-likelihood function of model \eqref{eq:mv-node-latent}--\eqref{eq:mv-node-subgroup-z} is computationally intractable. Thus, we will derive a multi-view log-pseudo-likelihood function for model \eqref{eq:mv-node-latent}--\eqref{eq:mv-node-subgroup-z}. We begin by approximating the conditional density of $\widehat b$ and $Y$ given $d$, where $\widehat b$ contains the number of edges connecting each of the $n$ nodes in the network to each of the $K$ estimated communities in the network, and $d$ contains the node degrees:
\begin{align} 
\label{eq:netapprox}
\widehat b, Y \mid \z[1], \z[2], d ~\dot \sim ~\prod \limits_{i=1}^n g(\widehat b_i; d_i, \eta_{\z[1]_i}) \phi(Y_i;  \gamma_{\z[2]_i}).
\end{align} 
The derivation of \eqref{eq:netapprox} is very similar to the derivation of \eqref{eq:approx2} in Section \ref{sec:multi-sbm-pseudo}. Ignoring any dependence between $d$ and $(\z[1], \z[2])$, and marginalizing over $\z[1]$ and $\z[2]$ in \eqref{eq:netapprox} to approximate the conditional distribution of $\widehat b$ and $Y$ given $d$, yields 
\begin{align} 
&\ell_{PL}(\eta, \gamma, \prob[1], \prob[2],  C; \widehat b, Y \mid d) =  \sum \limits_{i=1}^n \log \left ( \sum \limits_{k, k'} \prob[1]_k \prob[2]_{k'} C_{kk'} g(\widehat b_i; d_i, \eta_{k}) \phi(Y_i; \gamma_{k'}) \right ).\label{eq:node-pseudo}
\end{align} 
We observe that the  log-pseudo-likelihood function in \eqref{eq:node-pseudo} closely resembles \eqref{eq:pseudo-log-lik-joint}.

\subsection{Testing independence between $\z[1]$ and $\z[2]$} 
\label{sec:node-test}
We now propose a  test for the null hypothesis that the latent community memberships $\z[1]$ and the latent cluster memberships $\z[2]$ in model \eqref{eq:mv-node-latent}--\eqref{eq:mv-node-subgroup-z} are independent.  As in Section \ref{sec:multi-sbm-test}, this amounts to testing $H_0: C = 1_{\kk[1]} 1_{\kk[2]}^T$. 

Recall that the network $X$ marginally follows a SBM, and let $\widehat \eta$ and $\widehat \pi$ be the maximizers of $\ell_{PL}(\eta, \prob[1]; \widehat b \mid d)$, where $\ell_{PL}$ is the log-pseudo-likelihood function for the SBM given by \eqref{eq:pseudo-log-lik}. As in Section \ref{sec:test-stat}, we can compute $\widehat \eta$ and $\hprob[1]$ by using the EM algorithm for fitting FMMs \citep{mclachlan2007algorithm}. Recall that the rows of the multivariate view $Y$ marginally follow a FMM, and let $\widehat \gamma$ and $\hprob[2]$ be the maximizers of the log-likelihood function for the multivariate view, obtained via EM. We consider the $P^2$LRT statistic given by 
\begin{align*} 
\log  \widetilde \Lambda &\equiv \underset{C \in \mathcal{C}_{\hprob[1], \hprob[2]}}{\arg \max} ~ \ell_{PL}(\widehat \eta, \widehat \gamma, \hprob[1], \hprob[2], C; \widehat b, Y \mid d) - \ell_{PL}(\widehat \eta, \widehat \gamma, \hprob[1], \hprob[2], 1_{\kk[1]} 1_{\kk[2]}^T; \widehat b, Y \mid d), 
\end{align*} 
where $\ell_{PL}$ is the log-pseudo-likelihood function in \eqref{eq:node-pseudo}, and $\mathcal{C}_{\cdot, \cdot}$ is defined in Proposition \ref{prop:reparam}. Once again, we can perform the maximization over $C$ using techniques from convex optimization; details of the exponentiated gradient descent algorithm that we use are similar to Step 2 of Algorithm \ref{alg:est}.  As in Section \ref{sec:null-approx}, we approximate the null distribution of $\log \widetilde \Lambda$ by taking $M$ random permutations of the rows of $\x[2]$, and comparing the observed value of $\log \Lambda$ to its empirical distribution in the permuted data. Details are similar to Algorithm \ref{alg:test}.

\section{Related literature}
\label{sec:lit}
 Many papers have extended the SBM to the multiple network data view setting, under the assumption that a single set of communities is shared across all networks \citep{han2015consistent, peixoto2015inferring, paul2016consistent} or a subset of networks \citep{stanley2016clustering}. The model proposed in Section \ref{sec:multi-sbm-mod} does not rely on this assumption. Most of the previous work that avoids the assumption of shared communities has focused on estimation of the community structure; Section 4 of \citet{kim2018review} reviews these papers in detail. By contrast, the primary goal of our paper is not estimation, but rather to develop a test of association between the communities underlying each network view (Section \ref{sec:multi-sbm-test}). 

A related problem in functional neuroimaging is to test whether the communities underlying brain networks of two groups of healthy and diagnosed patients are the same; see \citet{paul2020random}, and the references contained therein. However, the test statistics and/or p-values for these tests cannot be computed in the two network data view setting. 

We proposed a test of the null hypothesis that the communities underlying two network views are independent. By contrast, \citet{xiong2019graph} proposed a test of the null hypothesis that the networks are \emph{conditionally} independent given their underlying communities.

In the case of a network view and a multivariate view, several papers have assumed that the communities underlying the network view and the clusters underlying the multivariate view are the same, and exploit this assumption to improve parameter estimation \citep{binkiewicz2017covariate, stanley2019stochastic, yan2020covariate}.  Our proposed model in Section \ref{sec:node-mod} does not rely on this assumption. Another body of work estimates the relationship between community memberships and node covariates, but does not consider inference on this relationship \citep{yang2013community,  newman2016structure, zhang2016community}. 

In Section \ref{sec:node}, we proposed testing for a specific type of relationship between the network view and the multivariate view: we test for association between the communities underlying the network view and the clusters underlying the multivariate view. Several papers have considered testing for other types of relationships between the network view and the multivariate view \citep{traud2011comparing, fosdick2015testing, peel2017ground}. For example, \citet{peel2017ground} tests for association between the network view and a categorical node covariate.

\section{Simulation results} 
\label{sec:sim}

In this section, we evaluate the power and Type I error of the tests proposed in Sections \ref{sec:multi-sbm-test}--\ref{sec:node}. 
Simulations in this paper were conducted using the \verb+simulator+ package \citep{bien2016simulator}.

\subsection{SBM for two network data views} 
\label{sec:networks-power}
We will evaluate the performance of four tests of $H_0: C = 1_{K^{(1)}}1_{K^{(2)}}^T$:
\begin{enumerate}[1.]
\item The $P^2$LRT proposed in Section \ref{sec:multi-sbm-test}, using the true values of $K^{(1)}$ and $K^{(2)}$,
\item The $P^2$LRT proposed in Section \ref{sec:multi-sbm-test}, using estimated values of $K^{(1)}$ and $K^{(2)}$,
\item The $G$-test for testing dependence between two categorical variables (Chapter 3.2, \citealt{agresti2003categorical}) applied to the estimated community memberships for each view, using the true values of $K^{(1)}$ and $K^{(2)}$, and
\item The $G$-test, using estimated values of $K^{(1)}$ and $K^{(2)}$.
\end{enumerate} 
We estimate $K^{(1)}$ and $K^{(2)}$ by applying the method of \citet{le2015estimating} to $\x[1]$ and $\x[2]$, respectively. In all four tests, we approximate the null distribution with a permutation approach, as in Algorithm \ref{alg:test}, using $M = 200$ permutation samples. 

\begin{figure}[h!]
\centering
\includegraphics[scale=0.5]{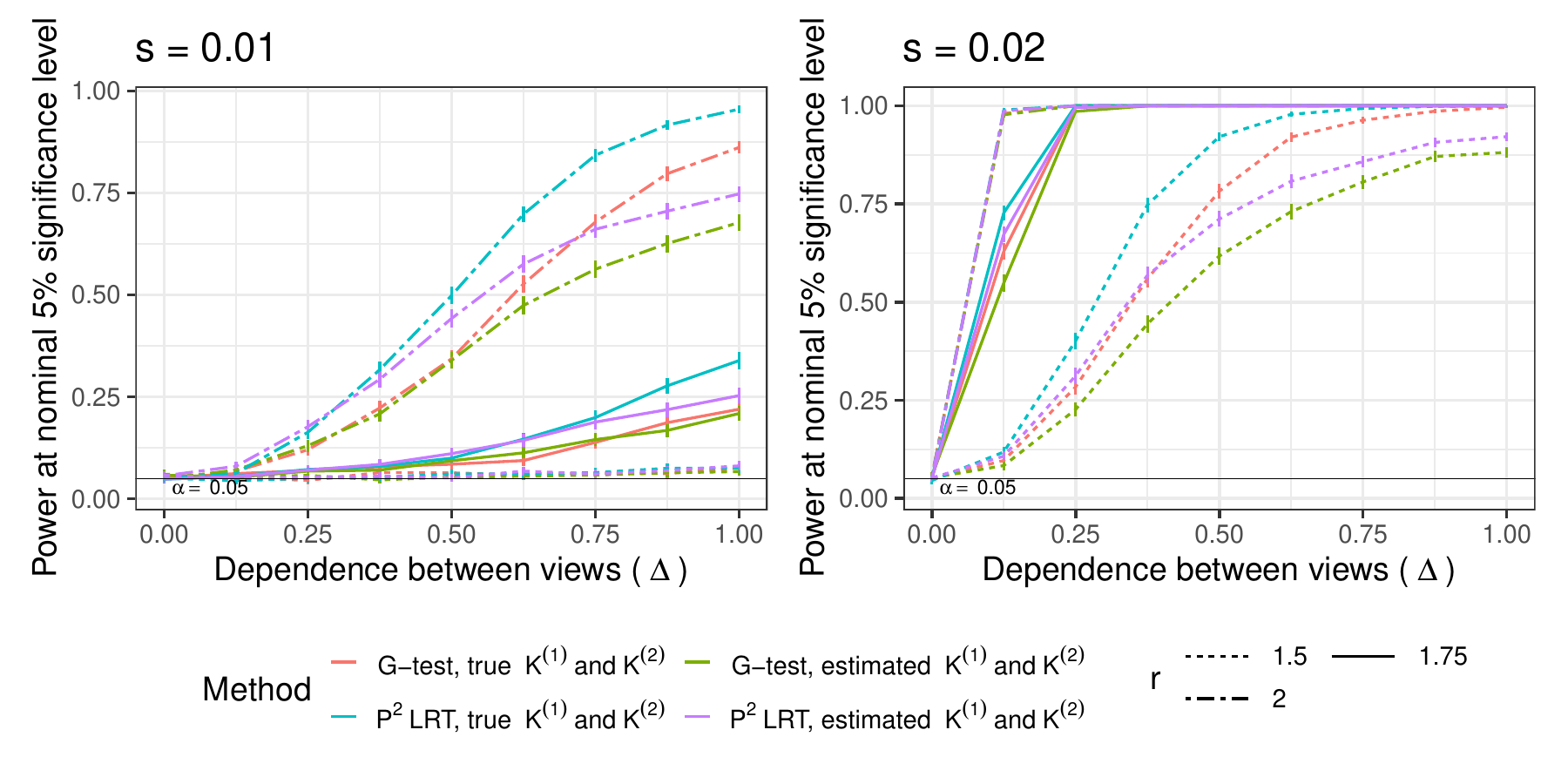}
\caption{\label{fig:sim1}
  Power of the $P^2$LRT and the $G$-test with both views drawn from a SBM, as we vary the dependence between views ($\Delta$), the strength of the communities ($r$), the expected edge density ($s$), and how the number of communities is selected. Details are in Section \ref{sec:networks-power}. 
}
\end{figure}

We generate data from model \eqref{eq:multi-sbm-latent} --\eqref{eq:mv-subgroup-z}, with $n = 1000$, $K^{(1)} = K^{(2)} = K = 6$, and 
\begin{align} 
C = (1 - \Delta) 1_K 1_K^T  + \Delta \cdot\text{diag}(K1_K),\label{eq:Csim} 
\end{align} 
for $\Delta \in [0, 1]$. Here, $\Delta = 0$ corresponds to independent communities and $\Delta = 1$ corresponds to identical communities. We let $\pi^{(1)} = \pi^{(2)} = 1_K/K$, and $\thet[1] = \thet[2] = \theta$, with
\begin{align} 
\theta_{kk'}= \omega (\mathds{1} \{k \neq k'\} + 2r \mathds{1} \{k = k' \}), \label{eq:thetsim} 
\end{align} 
for $r > 0$ describing the strength of the communities, and $\omega$ chosen so that the expected edge density of the network equals $s$, to be specified. We simulate 2000 data sets for a range of values of $s$, $\Delta$, and $r$, and evaluate the power of the four tests described above.  Results are shown in Figure \ref{fig:sim1}. 

 For all tests, power tends to increase as $\Delta$, which controls the dependence between views, 
increases. Power also tends to increase as the strength of the communities ($r$) increases, and as the expected edge density $(s)$ increases. Estimating $K^{(1)}$ and $K^{(2)}$ tends to yield lower power than using the true values of $K^{(1)}$ and $K^{(2)}$. All tests control the Type I error, but the $P^2$LRTs uniformly yield higher power than the $G$-tests. This is because the $P^2$LRT can be interpreted as a version of the $G$-test that replaces the ``hard" community assignments with ``soft" community assignments (Section 5, \citealt{gao2020clusterings}). Thus, the $P^2$LRT outperforms the $G$-test when the communities are more difficult to detect. 

We generate data with unbalanced community sizes in Appendix \ref{sec:supp-sim1}, and investigate how the true values of $\kk[1]$ and $\kk[2]$ relates to power in Appendix \ref{sec:supp-sim2}. 

\subsection{Degree-corrected SBM for two network data views} 
\label{sec:networks-dc-sim}

\begin{figure}[h!]
\centering
\includegraphics[scale=0.5]{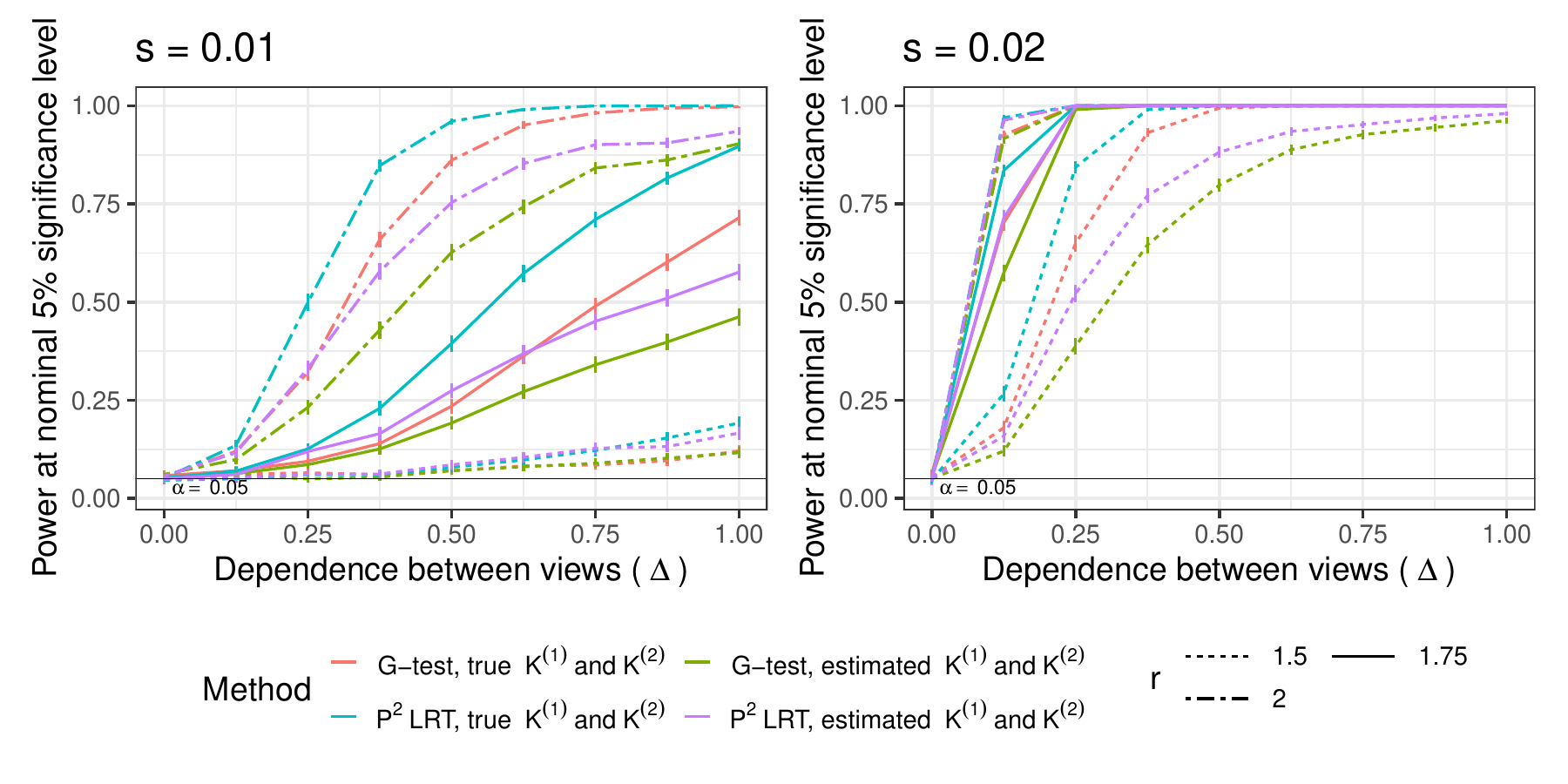}
\caption{\label{fig:sim2} Power of the $P^2$LRT and the $G$-test with both views drawn from a DCSBM, as we vary the dependence between views ($\Delta$), the strength of the communities ($r$), the expected edge density ($s$), and how the number of communities is selected. Details are in Section \ref{sec:networks-dc-sim}. }
\end{figure}

Under the SBM, nodes within the same community have the same expected degree. To investigate the performance of the test proposed in Section \ref{sec:multi-sbm-test} in a setting where nodes can have different expected degrees, we generate each network view from the degree-corrected stochastic block model (DCSBM, \citealt{karrer2011stochastic}). We generate $n$ vectors $(Z_i^{(1)}, Z_i^{(2)}, \delta_i^{(1)}, \delta_i^{(2)})$ i.i.d. for $i = 1, 2, \ldots, n$, with $\z[1]_i$ and $\z[2]_i$ categorical with $K^{(1)}$ and $K^{(2)}$ levels, respectively, and $(\z[1]_i, \z[2]_i) \perp (\del[1]_i, \del[2]_i)$. Here, $\delta^{(1)}$ and $\delta^{(2)}$ represent \emph{popularities} for the nodes in the two views; more popular nodes have higher expected degrees. We generate each view with
\begin{align} 
\x[l] \mid \z[l], \del[l] \sim \prod \limits_{j=1}^n \prod \limits_{i=1}^{j-1}  \left (\del[l]_i \del[l]_j\thet[l]_{Z_i Z_j} \right )^{\x[l]_{ij}} \left (1 -  \del[l]_i \del[l]_j\thet[l]_{\z[l]_i \z[l]_j} \right )^{1 - \x[l]_{ij}}, \quad l = 1, 2. \label{eq:dcsbm-latent-x}
\end{align}
We set $n$, $K^{(1)}$, $K^{(2)}$, $\pi^{(1)}, \pi^{(2)}$, $C$, $\theta^{(1)}$, and $\theta^{(2)}$ as in Section \ref{sec:networks-power} and take $\P(\del[l]_i = 2.5) = 0.2$, $\P(\del[l]_i = 0.625) = 0.8$, and $\del[1]_i \perp \del[2]_i$. We simulate 2000 data sets, varying the dependence between views $(\Delta)$, the expected edge density $(s)$, and the strength of the communities $(r)$; these parameters are defined in Section \ref{sec:networks-power}. Once again, we evaluate the power and Type I error of the four tests described in Section \ref{sec:networks-power}. Results are shown in Figure \ref{fig:sim2}, and are similar to Section \ref{sec:networks-power}. The $P^2$LRT performs well because it is based on an approximation to the conditional likelihood of the multi-view SBM given the node degrees (Section \ref{sec:multi-sbm-pseudo}); thus, it can handle the highly heterogeneous node degrees that characterize the multi-view DCSBM.

In this subsection, we assumed that the node popularities ($\del[1]$ and $\del[2]$) are independent. This can sometimes be an unrealistic assumption in practice. 
 If $\del[1]$ and $\del[2]$ are dependent, then $\x[1]$ and $\x[2]$ could be dependent even when the communities are independent, which could inflate the Type I error rate.  To investigate this effect, in Appendix \ref{sec:sim-setup}, we generate data from a multi-view DCSBM with $\del[1]$ and $\del[2]$ dependent, and apply the $P^2$LRT using a range of values of $K^{(1)}$ and $K^{(2)}$. We find that the Type I error rate is controlled, both when we estimate the number of communities and when we choose a fixed number of communities (as long as the number of communities is not grossly overspecified); Appendix \ref{sec:explain} gives intuition for why this is the case.

\subsection{SBM for a network view and a multivariate view} 
\label{sec:node-sim}
We will evaluate the performance of six tests of $H_0: C = 1_{K^{(1)}}1_{K^{(2)}}^T$:
\begin{enumerate}[1.] 
\item The $P^2$LRT proposed in Section \ref{sec:node}, using the true values of $K^{(1)}$ and $K^{(2)}$,
\item The $P^2$LRT, using estimated values of $K^{(1)}$ and $K^{(2)}$,
\item The $G$-test applied to the estimated community/cluster memberships in the network/multivariate view, using the true values of $K^{(1)}$ and $K^{(2)}$,
\item The $G$-test, using estimated values of $K^{(1)}$ and $K^{(2)}$,
\item The BEStest \citep{peel2017ground} applied to the network view and the estimated cluster memberships in the multivariate view, using the true values of $K^{(1)}$ and $K^{(2)}$, 
\item The BESTest, using estimated values of $K^{(1)}$ and $K^{(2)}$.
\end{enumerate} 
We estimate $K^{(1)}$ by applying the method of \citet{le2015estimating}, we estimate $K^{(2)}$ using BIC, and we approximate the null distributions using $M = 200$ permutation samples.  

\begin{figure}[h!]
\centering
\includegraphics[scale=0.4]{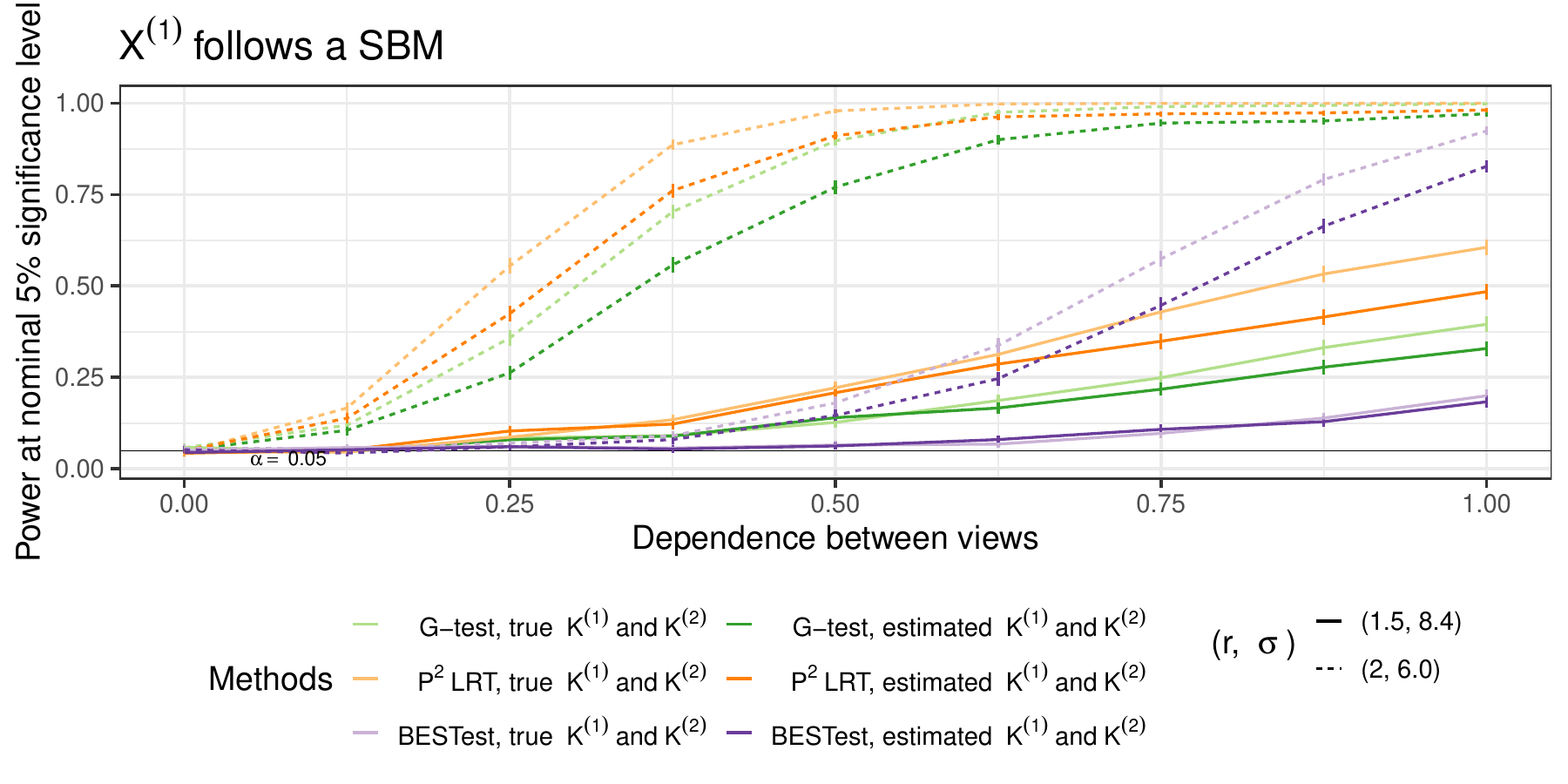}
\caption{\label{fig:sim3} Power of the $P^2$LRT, the $G$-test, and the BESTest \citep{peel2017ground} with the multivariate view drawn from a Gaussian mixture model and the network view drawn from a SBM, as we vary the dependence between views ($\Delta$), the strength of the communities ($r$), the variance of the clusters $(\sigma$), and how the number of communities and the number of clusters are selected. The expected edge density $(s)$ is fixed at $0.015$. Details are in Section \ref{sec:node-sim}. }
\end{figure}

We generate data from model \eqref{eq:mv-node-latent}--\eqref{eq:mv-node-subgroup-z}; we generate data from a degree-corrected version of model \eqref{eq:mv-node-latent}--\eqref{eq:mv-node-subgroup-z} in Appendix \ref{sec:dcnode}. We set $n = 500$, and $K^{(1)} = K^{(2)} = K = 3$. Let $\pi^{(1)} =  \pi^{(2)} = 1_K/K$, and let $C$ be given by \eqref{eq:Csim}. Let $\theta$ be given by \eqref{eq:thetsim}, so that the expected edge density is $s = 0.015$. We draw the multivariate data view from a Gaussian mixture model, for which the $k$th mixture component is a $N_{10}(\mu_k, \sigma^2 I_{10})$ distribution. The $p \times K$ mean matrix for the multivariate data view is given by $\mu  = \left [ \begin{matrix} 0 \cdot 1_5 & 0 \cdot 1_5 & \sqrt{12} \cdot 1_5 \\ 2 \cdot 1_5 & -2 \cdot 1_5 & 0 \cdot 1_5 \end{matrix} \right ]$. We simulate 2000 data sets for a range of values of $\Delta$, $r$, and $\sigma$. Results are shown in Figure \ref{fig:sim3}.

All tests control the Type I error rate. Power tends to increase as the dependence between views ($\Delta$) increases. Power also tends to increase as the strength of the communities ($r$) increases and the variance of the clusters ($\sigma$)  decreases. The $P^2$LRTs uniformly yield higher power than the $G$-tests and the BESTests.

\section{Application to protein-protein interaction data}
\label{sec:app}
In this section, we focus on two types of protein-protein interaction data. A binary interaction is a physical interaction between proteins, and a co-complex association is a pair of proteins that are part of the same complex. These two data views represent distinct biological concepts; physical interactions can occur between a pair of proteins that are not in the same complex, and not all proteins in complexes physically interact. 

To investigate whether the latent communities of proteins defined with respect to binary interactions and co-complex associations are related, we consider \emph{H. sapiens} protein-protein interaction data from the HINT (High-quality INteractomes; \citet{das2012hint}) database, and ask: are the communities within the binary network and the communities within the co-complex network associated? 

\begin{figure}[h!] 
\centering 
\includegraphics[scale=0.5]{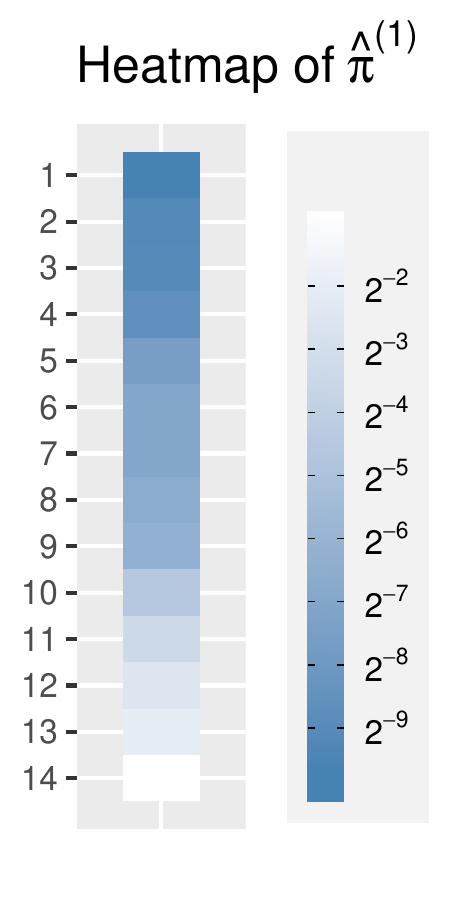} \includegraphics[scale=0.5]{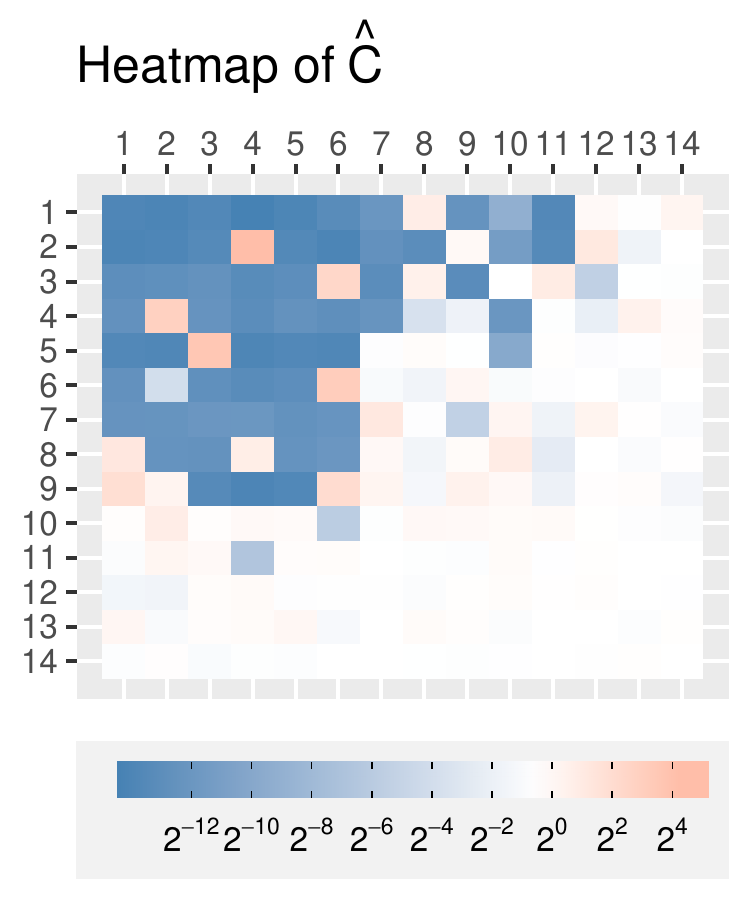} \includegraphics[scale=0.5]{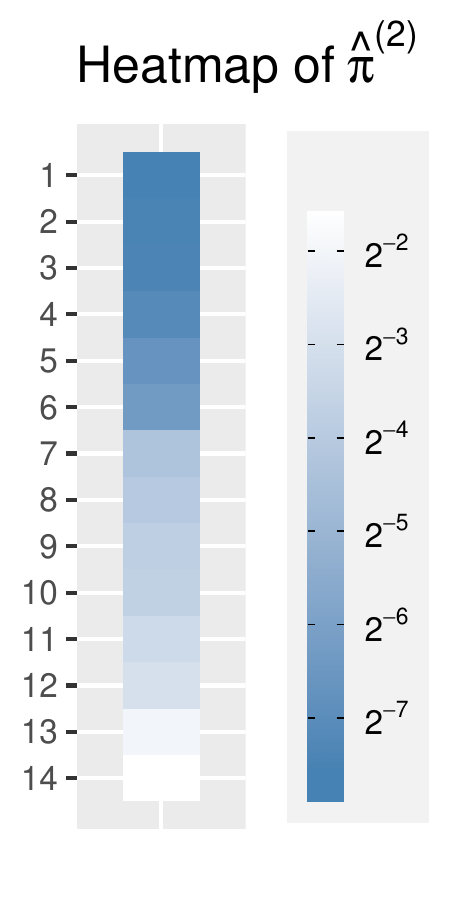}
\caption{\label{fig:Chat}  Heatmaps of $\hprob[1]$ and $\hprob[2]$, defined in Section \ref{sec:test-stat}, and of $\widehat C$, defined in \eqref{eq:pseudoCopt}, for the HINT data described in Section \ref{sec:app}. } 
\end{figure}

We remove self-interactions from both networks, and consider
only those proteins that appear in both networks.
This yields $43,874$ binary interactions and $88,960$ co-complex associations among a common set of $n = 9,037$ proteins. We apply the $P^2$LRT of $H_0: C = 1_{K^{(1)}} 1_{K^{(2)}}^T$ developed in Section \ref{sec:multi-sbm-test}, using $M = 10^4$ in Step 3 of Algorithm \ref{alg:test}. As in Section \ref{sec:sim}, we estimate the number of communities in each view by applying the method of \citet{le2015estimating} to each view separately, 
which (coincidentally) estimates 14 communities in both data views. Figure \ref{fig:Chat} displays $\widehat \pi^{(1)}$ and $\widehat \pi^{(2)}$ (defined in Section \ref{sec:test-stat}), and $\widehat C$ (defined in equation \ref{eq:pseudoCopt}). Our test yields a p-value of $0.013$, and thus provides some evidence against the null hypothesis  that communities of proteins defined with respect to binary interactions and communities of proteins defined with respect to co-complex associations are independent.

Our test of $H_0: C = 1_{\kk[1]} 1_{\kk[2]}^T$ allows us to provide an answer to the high-level scientific question of whether there is a relationship between communities defined with respect to different types of protein interactions. However, it may also be of scientific interest to determine whether there is a relationship between the $k$th community in the binary view and the $k'$th community in the co-complex view. Recall from Section \ref{sec:multi-sbm-mod} that $C_{kk'} = 1$ indicates that the $k$th community in the binary view and the $k'$th community in the co-complex view are independent. In Figure \ref{fig:Chat}, most values of $\widehat C_{kk'}$ are close to 1. Thus, it may be of future interest to develop tests of $H_0: C_{kk'} = 1$.

\section{Discussion}
\label{sec:discuss}

In this paper, we considered testing whether communities defined with respect to two networks on a common set of nodes are related. We extended this test to the setting of one network and one multivariate data set on a common set of nodes. The proposed tests control the Type I error rate, and yield higher power than applying the $G$-test to the estimated community/cluster memberships in each data view.

We focused on testing the association between communities/clusters in two data views. If three or more data views are available, we may be interested in testing mutual independence between all data views. The models proposed in Sections \ref{sec:multi-sbm-mod} and \ref{sec:node-mod} extend readily to $L > 2$ data views, and we can test for mututal independence by testing the null hypothesis that all entries of an $L$th order tensor $C$ are equal to 1. We can construct a $P^2$LRT statistic along the lines of \eqref{eq:genp2lrt}, and we can approximate the null distribution by permuting the node labels in the second through $L$th views. If we are instead interested in pairwise independence between the data views, we could simply apply the tests developed in this paper to each pair of views. 

In this paper, we considered only undirected, unweighted network views. There is a body of work that extends the single-view SBM to directed and/or weighted networks; see e.g. \citet{wang1987stochastic} and \citet{aicher2014learning}. It may be of future interest to extend the methodology developed in this paper to allow for directed and/or weighted networks.

\section*{Acknowledgments}
Lucy L. Gao received funding from the Natural Sciences and Engineering Research Council of Canada. Daniela Witten and Jacob Bien were supported by NIH Grant R01GM123993.  Jacob Bien was supported by NSF CAREER Award DMS-1653017. Daniela Witten was supported by NIH Grant DP5OD009145, NSF CAREER Award DMS-1252624, and Simons Investigator Award No. 560585. We thank Haiyuan Yu for useful input on protein interaction data.
{\it Conflict of Interest}: None declared.

\section*{Data Avalability Statement} 
The data that support the findings of this paper are openly available in the HINT (High-quality INTeractions) database at \verb+http://hint.yulab.org+ \citep{hint}.

\section*{Supporting Information} 
  The tests developed in this paper are implemented in the R package \verb+multiviewtest+, which is available on CRAN. Code to reproduce the results in this paper is available at \\  \verb+https://github.com/lucylgao/mv-network-test-code+.

\bibliographystyle{biom} \bibliography{refs}

\appendix

\section{A detailed review of \citet{amini2013pseudo}}
\label{sec:approx}

Let $\widehat Z \in\{1, \ldots, K\}^n$ be an initial estimate of the community memberships of the $n$ nodes. Specifically, \citet{amini2013pseudo} proposed using a regularized spectral clustering procedure called spectral clustering with perturbations to obtain $\widehat Z$.  In what follows, the dependency of $\widehat{Z}$ on $X$ is ignored, and $\widehat{Z}$ is treated as fixed. Let $\widehat b$ be the $n \times K$ matrix defined by
\begin{align} 
\widehat{b}_{im} = \sum \limits_{j=1}^n X_{ij} \mathds{1}\{\widehat{Z}_j = m\}, \quad 1 \leq i \leq n, 1 \leq m \leq K. \label{eq:block-row-sums}
\end{align}
Let $\widehat b_i$ denote the $i$th row of $\widehat b$. Let $d = X 1_n$. In this section, we review the derivation of a pseudolikelihood function from \citet{amini2013pseudo} which is based on an approximation to the conditional density of $\widehat b$ given $d$. We note that \citet{amini2013pseudo} also derived a pseudolikelihood function which is based on the unconditional density of $\widehat b$. However, the estimators which maximize the former pseudolikelihood function are more robust against misspecification of the conditional distribution of $X$ given $Z$ in the stochastic block model (Section \ref{sec:sbm-mod}) than the estimators which maximize the latter pseudolikelihood function \citep{amini2013pseudo}. This is because the conditional distribution of $X$ given $Z$ in  the stochastic block model provides a poor fit to networks with heterogeneous node degrees within communities, and conditioning on $d$ (the node degrees) improves the goodness of fit. 

It follows from the definition of the stochastic block model (Section \ref{sec:sbm-mod}) that:
\begin{itemize} 
\item For $(i, j), (i', j') \in \{1, 2, \ldots, n\}^2$, conditional on $Z$, $X_{ij} \perp X_{i'j'}$, and 
\item For $(i, j, m), (i', j', m') \in \{1, 2, \ldots, n\} \times \{1, 2, \ldots, n\} \times \{1, 2, \ldots, K\}$, conditional on $Z$, 
\begin{align} 
X_{ij} \mathds{1}\{\widehat{Z}_j = m\} \perp X_{i'j'} \mathds{1}\{\widehat{Z}_{j'} = m'\}. \label{eq:condind}
\end{align} 
\end{itemize} 
Thus, conditional on $Z$, $\{(\widehat b_i, d_i)\}_{i=1}^n$ are weakly dependent when $n$ is large, and so
\begin{align} 
\label{eq:prod}
f(\{\widehat b_i \}_{i=1}^n \mid Z, d) = &\frac{f(\{\widehat b_i \}_{i=1}^n, d \mid Z)}{f(d \mid Z)} 
\approx \frac{\prod \limits_{i=1}^n f(\widehat b_i, d_i \mid Z)}{\prod \limits_{i=1}^n f(d_i \mid Z)} = \prod \limits_{i=1}^n f(\widehat b_i \mid Z, d_i).
\end{align} 

Next, we derive approximations to $f(\widehat b_i \mid Z, d_i)$. Recall from the definition of the stochastic block model (Section \ref{sec:sbm-mod}) that conditional on $Z$, $X_{ij}$ are independent Bernoulli variables for $1 \leq i < j \leq n$. Thus, it follows from the definition of $\widehat b_{im}$ in \eqref{eq:block-row-sums} that conditional on $Z$, $\widehat b_{im}$ is the sum of independent Bernoulli random variables, and can be approximated by a Poisson distribution: 
\begin{align} 
\label{eq:poissonapprox} 
\widehat b_{im} \mid Z ~\dot \sim~ \text{Poisson}\left (\sum \limits_{j=1}^n \mathbb{E}[X_{ij} \mathds{1} \{\widehat Z_j = m\} \mid Z] \right ). 
\end{align} 
Ignoring the fact that $X_{ii} = 0$, and instead assuming that $X_{ii} \mid Z \sim $ Bernoulli$(\theta_{Z_i Z_i})$ with $\{X_{ij} \}_{1 \leq j \leq i \leq n}$ conditionally independent given $Z$,
\begin{align} 
\label{eq:poissonmean} 
\mathbb{E}[\widehat{b}_{im} \mid Z] \approx \sum \limits_{j=1}^n \theta_{Z_i Z_j} \mathds{1} \{ \widehat{Z}_j = m\} = \sum \limits_{j=1}^n \sum \limits_{m' = 1}^K \theta_{Z_i m'} \mathds{1} \{\widehat{Z}_j = m, Z_j = m' \} =  \sum \limits_{m'=1}^K \theta_{Z_i m'} \widehat R_{mm'}, 
\end{align} 
where $\widehat{R}$ is the confusion matrix of $\widehat{Z}$ defined by 
\begin{align} 
\label{eq:confusion}
\widehat{R}_{mm'} = \sum \limits_{j=1}^n  \mathds{1} \{\widehat{Z}_j = m, Z_j = m' \}, \quad 1 \leq m \leq K, 1 \leq m' \leq K.
\end{align}
Combining \eqref{eq:poissonapprox} and \eqref{eq:poissonmean}, 
\begin{align} 
\widehat{b}_{im} \mid  Z ~\dot \sim ~\mathrm{Poisson} \left (\sum \limits_{m'=1}^K \theta_{Z_i m'} \widehat R_{mm'} \right ), \quad 1 \leq i \leq n, 1 \leq m \leq K.  \label{eq:poisson}
\end{align} 
Now, the joint distribution of independent Poisson random variables conditional on their sum is multinomial. It follows from \eqref{eq:block-row-sums} and \eqref{eq:condind} that $\{\widehat b_{im} \}_{i=1}^n$ are conditionally independent given $Z$. Furthermore, from \eqref{eq:poisson}, conditional on $Z$, $\widehat b_{im}$ are approximately Poisson. Thus, 
\begin{align} 
\label{eq:multi}
\widehat{b}_i \mid d_i, Z ~\dot \sim ~\mathrm{Multinomial}\left (d_i,   \left (\frac{\sum \limits_{m'=1}^K \theta_{Z_i m'} \widehat R_{1m'}}{\sum \limits_{m=1}^K \sum \limits_{m'=1}^K \theta_{Z_i  m'} \widehat R_{mm'}}, 
\ldots, \frac{\sum \limits_{m'=1}^K \theta_{Z_i m'} \widehat R_{Km'}}{\sum \limits_{m=1}^K \sum \limits_{m'=1}^K \theta_{Z_i  m'} \widehat R_{mm'}}, \right ) \right ), \quad 1 \leq i \leq n. 
\end{align}
We use \eqref{eq:multi} to write 
\begin{align} 
\label{eq:multi-like-paper} 
\widehat{b}_i \mid d_i, Z ~\dot \sim ~ g(\widehat b_i; d_i, \eta_{Z_i}), \quad 1 \leq i \leq n,
\end{align} 
where $g(\cdot; q)$ denotes the probability mass function of a Multinomial$(N, q_1, \ldots, q_K)$ random variable, and $\eta = \left (\text{diag}(\theta \widehat R1_K) \right )^{-1} \theta \widehat R$. 
Now, combining \eqref{eq:prod} and \eqref{eq:multi-like-paper}, 
\begin{align} 
\widehat b \mid Z, d  ~\dot \sim ~  \prod \limits_{i=1}^n g(\widehat b_i; d_i, \eta_{Z_i}). \label{eq:prodapprox}
\end{align} 
Treating $\eta$ as fixed, and  marginalizing over $Z$ in \eqref{eq:prodapprox}, ignoring any dependency of $d$ on $Z$, yields
\begin{align} 
\widehat b \mid d ~ \dot \sim ~\prod \limits_{i=1}^n \left ( \sum \limits_{k=1}^K \pi_k  g(\widehat b_i; d_i, \eta_k) \right ). \label{eq:bapprox}
\end{align} 
Based on \eqref{eq:bapprox}, \citet{amini2013pseudo} defined the log-pseudolikelihood function to be:
\begin{align*} 
\ell_{PL}(\eta, \pi) = \sum \limits_{i=1}^n \log \left ( \sum \limits_{k=1}^K \pi_k g(\widehat{b}_i; d_i,  \eta_k) \right ).
\end{align*} 
This is \eqref{eq:pseudo-log-lik}.

\section{The DCSBM for two network data views with dependent popularities}
\label{sec:more-sim}
In Section \ref{sec:networks-dc-sim}, we generated data from a DCSBM for two network data views, where $\del[1]$ (the popularities of the nodes in the first view) and $\del[2]$ (the popularities of the nodes in the second view) are independent. In this section, we will modify the DCSBM for two network data views to a case of maximal dependence between the node popularities of the two views: $\del[1]_i = \del[2]_i$ for all $i = 1, 2, \ldots, n$. 

\subsection{Type I error rate of the $P^2$LRT} 
\label{sec:sim-setup}
We will generate each network view from the DCSBM. We generate $n$ vectors $(Z_i^{(1)}, Z_i^{(2)}, \delta_i^{(1)}, \delta_i^{(2)})$ i.i.d. for $i = 1, 2, \ldots, n$, with $\z[1]_i$ and $\z[2]_i$ categorical with $K^{(1)}$ and $K^{(2)}$ levels, respectively, and $(\z[1]_i, \z[2]_i) \perp (\del[1]_i, \del[2]_i)$. We let $\del[1]_i = \del[2]_i$ for $i = 1, 2, \ldots, n$, so that the node popularities in the two views are identical. We generate each view with
\begin{align*} 
\x[l] \mid \z[l], \del[l] \sim \prod \limits_{j=1}^n \prod \limits_{i=1}^{j-1}  \left (\del[l]_i \del[l]_j\thet[l]_{\z[l]_i \z[l]_j} \right )^{\x[l]_{ij}} \left (1 -  \del[l]_i \del[l]_j\thet[l]_{\z[l]_i \z[l]_j} \right )^{1 - \x[l]_{ij}}, \quad l = 1, 2.
\end{align*}
We set $n = 50$, $K^{(1)} = K^{(2)} = K = 2$, $\prob[1] = \prob[2] = 1_2/2$, $\thet[1] = \thet[2] = \left [ \begin{matrix} 0.5 & 0.25 \\ 0.25 & 1 \end{matrix} \right ]$, and $\del[1]_i \sim$ Uniform$(0.14, 0.84)$. We let $C = 1_2 1_2^T$, so that $\z[1]$ and $\z[2]$ are independent. We simulate 200 data sets with $C = 1_21 _2^T$. 

\begin{figure}[h!]
\centering
\includegraphics[scale=0.5]{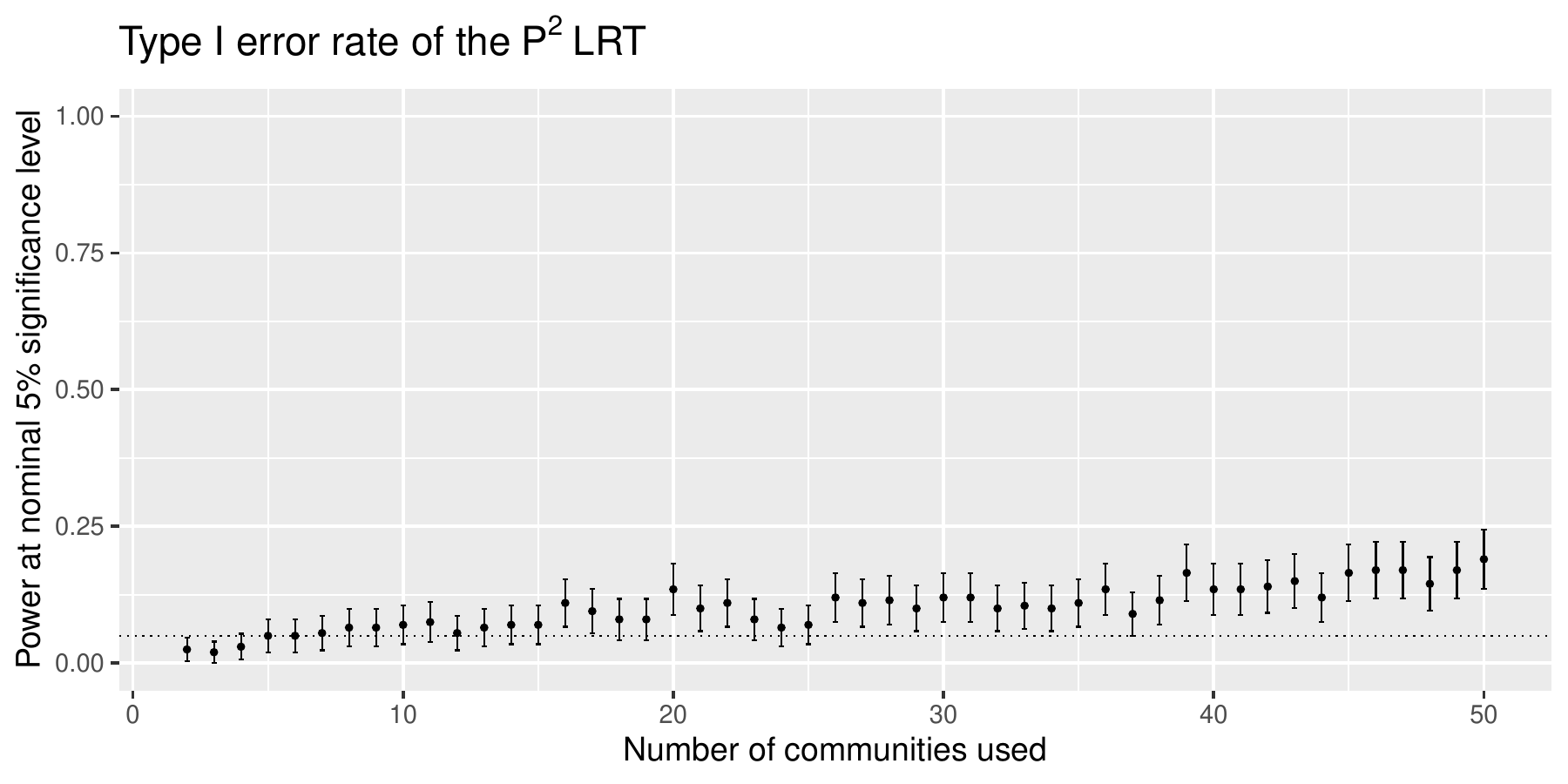}
\caption{\label{fig:asim1} For the simulation study described in Appendix \ref{sec:sim-setup}, we display the Type I error rate of the $P^2$LRT described in Section \ref{sec:multi-sbm-test} for $n = 50$, $K = 2$, and $\del[1]_i = \del[2]_i$ for $i = 1, 2, \ldots, n$. The x-axis displays the number of communities used, and the y-axis displays Type I error rate. The Type I error rate of the $P^2$LRT with the value of $K^{(1)}$ and $K^{(2)}$ estimated  by applying the method of \citet{le2015estimating} to $\x[1]$ and $\x[2]$, respectively, is $0.025$ (95\% confidence interval: $0.0034, 0.0466)$.}
\end{figure}

We apply the $P^2$LRT of $H_0: C = 1_{K^{(1)}} 1_{K^{(2)}}^T$ described in Section \ref{sec:multi-sbm-test}, using the same number of communities in each data view, and varying the number of communities used from 2 to $n = 50$. We also apply the $P^2$LRT using the value of $K^{(1)}$ and $K^{(2)}$ estimated by applying the method of \citet{le2015estimating} to $\x[1]$ and $\x[2]$, respectively. The results are shown in Figure \ref{fig:asim1}. 

We see that when we grossly overspecify the number of communities, the Type I error rate is inflated, and when we do not grossly overspecify the number of communities, the Type I error rate is controlled at the nominal $\alpha = 0.05$ level. 

\subsection{Number of communities used and Type I error rate } 
\label{sec:explain}
In this subsection, we will explain why the Type I error rate is inflated when $\del[1]$ and $\del[2]$ are dependent and we grossly overspecify the number of communities. 

The $P^2$LRT statistic defined in \eqref{eq:genp2lrt} is closely related to the mutual information (a measure of dependence, \citealt{meilua2007comparing}) between the estimated community memberships in each view; the derivation of this relationship is similar to Section 5 of \citet{gao2020clusterings}. This suggests that if the community memberships in the two views are independent, but the \emph{estimated} community memberships in the two views are dependent, then the Type I error rate will be inflated. Furthermore, if
\begin{enumerate}[1.]
\item the estimated community assignments in view 1 and $\del[1]$ are dependent, 
\item  the estimated community assignments in view 2 and $\del[2]$ are dependent, and 
\item $\del[1]$ and $\del[2]$ are dependent, 
\end{enumerate} 
then the estimated community assignments in the two views will likely be dependent. 

In Appendix \ref{sec:sim-setup}, we generate data with $\del[1]$ and $\del[2]$  dependent. When we specify a very large number of communities, the estimation procedure tends to assign nodes with similar values of $\del[l]$ to the same community. Thus, Conditions 1--3 above are satisfied, leading to dependence between the estimated community memberships, and hence Type I error inflation. 

When we do not grossly overspecify the number of communities, the estimated community assignments are not highly dependent on $\del[l]$, and thus the $P^2$LRT controls the Type I error rate. Estimating the number of communities using the method of \citet{le2015estimating} controls the Type I error rate, because the method of \citet{le2015estimating} does not grossly overspecify the number of communities. 

\section{DCSBM for a network view and a multivariate view}
\label{sec:dcnode}
We will evaluate the performance of four tests of $H_0: C = 1_{K^{(1)}}1_{K^{(2)}}^T$:
\begin{enumerate}[1.] 
\item The $P^2$LRT proposed in Section \ref{sec:node}, using the true values of $K^{(1)}$ and $K^{(2)}$ ,
\item The $P^2$LRT, using estimated values of $K^{(1)}$ and $K^{(2)}$,
\item The $G$-test applied to the estimated community assignments in the network view and the estimated cluster memberships in the multivariate view, using the true value of $K^{(1)}$ and $K^{(2)}$, and 
\item The $G$-test, using the estimated values of $K^{(1)}$ and $K^{(2)}$,
\end{enumerate} 
where $K^{(1)}$ (the number of communities in the network view) is estimated by applying the method of \citet{le2015estimating} to $X$, and $K^{(2)}$ (the number of clusters in the multivariate view) is estimated using BIC. In all four tests, we approximate the null distribution with a permutation approach, as in Algorithm \ref{alg:test}, using $M = 200$ permutation samples.  

We generate the network data view from a DCSBM, and the multivariate data view from a Gaussian mixture model. We generate $n$ vectors $(Z_i^{(1)}, Z_i^{(2)}, \delta_i)$ i.i.d. for $i = 1, 2, \ldots, n$, with $\z[1]_i$ and $\z[2]_i$ categorical with $K^{(1)}$ and $K^{(2)}$ levels, respectively, and $(\z[1]_i, \z[2]_i) \perp \delta_i$. We generate the network view with
\begin{align*} 
X \mid \z[1], \delta \sim \prod \limits_{j=1}^n \prod \limits_{i=1}^{j-1}  \left (\delta_i \delta_j\theta_{\z[1]_i \z[1]_j} \right )^{X_{ij}} \left (1 -  \delta_i \delta_j\theta_{\z[1]_i \z[1]_j} \right )^{1 - X_{ij}},
\end{align*}
and generate the multivariate data view with 
\begin{align*} 
Y \mid \z[2] \sim \prod \limits_{i=1}^n \phi(Y_i; \mu_k, \sigma^2 I_{10}), 
\end{align*} 
where $\phi(\cdot; \mu, \Sigma)$ denotes the density of a $N_{10}(\mu, \Sigma)$ random variable. The mean matrix for the multivariate data view is given by $\mu  = \left [ \begin{matrix} 0 \cdot 1_5 & 0 \cdot 1_5 & \sqrt{12} \cdot 1_5 \\ 2 \cdot 1_5 & -2 \cdot 1_5 & 0 \cdot 1_5 \end{matrix} \right ]$.

\begin{figure}[h!]
\centering
\includegraphics[scale=0.55]{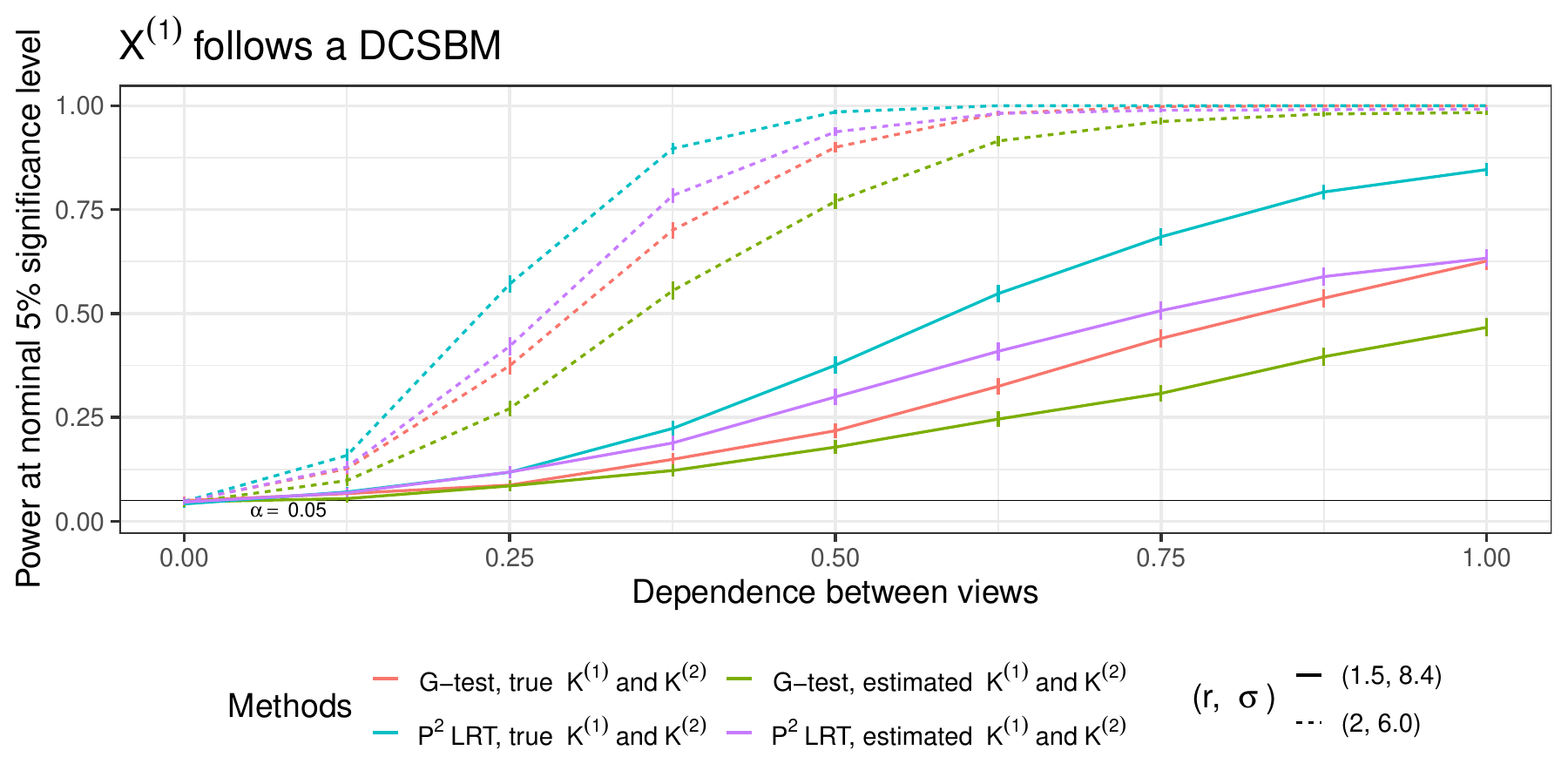}
\caption{\label{fig:dcnodesim} Power of the $P^2$LRT and the $G$-test with the multivariate view drawn from a Gaussian mixture model and the network view drawn from a DCSBM, varying the dependence between views ($\Delta$), the strength of the communities ($r$), the variance of the clusters $(\sigma$), and how the number of communities and the number of clusters are selected. The expected network density $(s)$ is fixed at $0.015$. Details are in Appendix \ref{sec:dcnode}.}
\end{figure}

We set $n = 500$, and $K^{(1)} = K^{(2)} = K = 3$. Let $\pi^{(1)} = \pi^{(2)} = 1_K/K$, and let $C$ be given by \eqref{eq:Csim}. Let $\theta$ be given by \eqref{eq:thetsim}, so that the expected edge density $s$ equals $0.015$. We simulate 2000 data sets for $n = 500$ and a range of values of $\Delta$, $r$, and $\sigma$. Results are shown in Figure \ref{fig:dcnodesim}, and are similar to the results in Section \ref{sec:node-sim}.

\section{SBM for two network data views with unbalanced community sizes}
\label{sec:supp-sim1} 
In Section \ref{sec:networks-power}, we generated data from model \eqref{eq:multi-sbm-latent} --\eqref{eq:mv-subgroup-z} with $\pi^{(1)} = \pi^{(2)} = 1_K/K$, so that the community sizes are balanced in the two views. In this section, we will instead let $\prob[1] = \prob[2] = (0.05, 0.05, 0.15, 0.15, 0.3, 0.3)^T$, so that the community sizes are unbalanced. Let $n = 1000$, $K^{(1)} = K^{(2)} = K = 6$, and let $C$ be given by
$$ C = (1 - \Delta) 1_K 1_K^T + \Delta \left [ \begin{matrix} 8 & 0 & 0 & 0 & 1 &  1 \\ 0 & 8 & 0 & 0 & 1 &  1\\ 0 & 0 & 8/3 & 0 & 1 & 1 \\ 0 & 0 & 0 & 8/3 & 1 & 1\\ 1 & 1 & 1 & 1 & 1 & 1 \\ 1 & 1 & 1 & 1 & 1 & 1\end{matrix} \right ], $$
for $\Delta \in [0, 1]$, so that any dependence between views comes from the smallest four communities in each view.
 Furthermore, instead of setting $\thet[1] = \thet[2]$ as in Section \ref{sec:networks-power}, we set
\begin{align*} 
\thet[1]_{kk'}&= \omega \left ( 2r \mathds{1} \{k = k'\}  +  \mathds{1} \{k \neq k' \} \right ), \\
\thet[2]_{kk'}&= \omega \left ( \mathds{1} \{k = k'\}  +  2r\mathds{1} \{k \neq k' \} \right ), 
\end{align*} 
for $\omega$ chosen so that the expected edge density $(s)$ of the network equals $0.025$, and $r > 1$ describing the strength of the communities. We simulate 2000 data sets for a range of values of $r$ and $\Delta$ (defined in \eqref{eq:Csim}), and evaluate the power of the four tests described in Section \ref{sec:networks-power}. 

\begin{figure}[h!]
\centering
\includegraphics[scale=0.55]{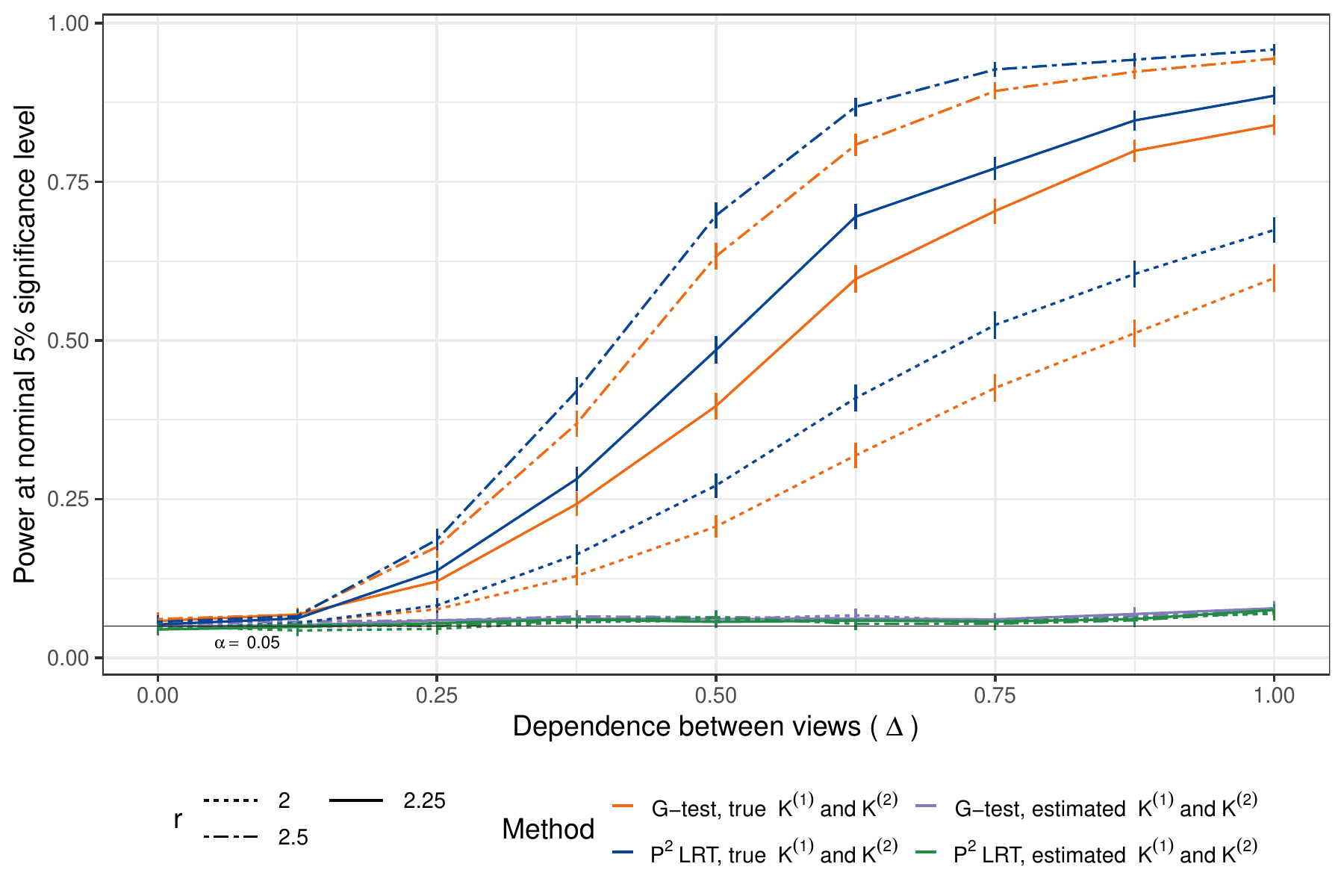}
\caption{\label{fig:s3}  Power of the $P^2$LRT and the $G$-test with both views drawn from a SBM, varying the dependence between views ($\Delta$), the strength of the communities ($r$), and how the number of communities is selected. The expected network density $(s)$ is fixed at $0.025$. Details are in Appendix \ref{sec:supp-sim1}.}
\end{figure} 

In  Figure \ref{fig:s3}, the $P^2$LRT still achieves uniformly higher power than the $G$-test when we use the true number of communities in each view. However, both the $G$-test and the $P^2$LRT have extremely low power when the number of communities in each view is estimated. This is because we tend to estimate $\kk[1] = \kk[2] = 3$, and the natural way to cluster each data view into three communities is to combine the four smallest communities into one ``meta-community". Since the dependence between views comes from the four smallest communities, when we combine them into one ``meta-community", there is little to no dependence between views left. Thus, whether we use the $G$-test or our proposed $P^2$LRT, testing for association using just three communities in each view yields extremely low power. 

\section{The effect of increasing the number of communities}
\label{sec:supp-sim2} 
In this section, we will fix the number of nodes ($n$), and investigate the performance of the test proposed in Section \ref{sec:multi-sbm-test} as a function of the number of communities in each view. We will evaluate the performance of two tests of $H_0: C = 1_{K^{(1)}}1_{K^{(2)}}^T$:
\begin{enumerate}[1.]
\item The $P^2$LRT proposed in Section \ref{sec:multi-sbm-test}, using the true values of $K^{(1)}$ and $K^{(2)}$,
\item The $G$-test for testing dependence between two categorical variables (Chapter 3.2, \citealt{agresti2003categorical}) applied to the estimated community assignments for each view, using the true values of $K^{(1)}$ and $K^{(2)}$.
\end{enumerate} 
In both tests, we approximate the null distribution with a permutation approach, as in Algorithm \ref{alg:test}, using $M = 200$ permutation samples. 

\begin{figure}[h!]
\centering
\includegraphics[scale=0.55]{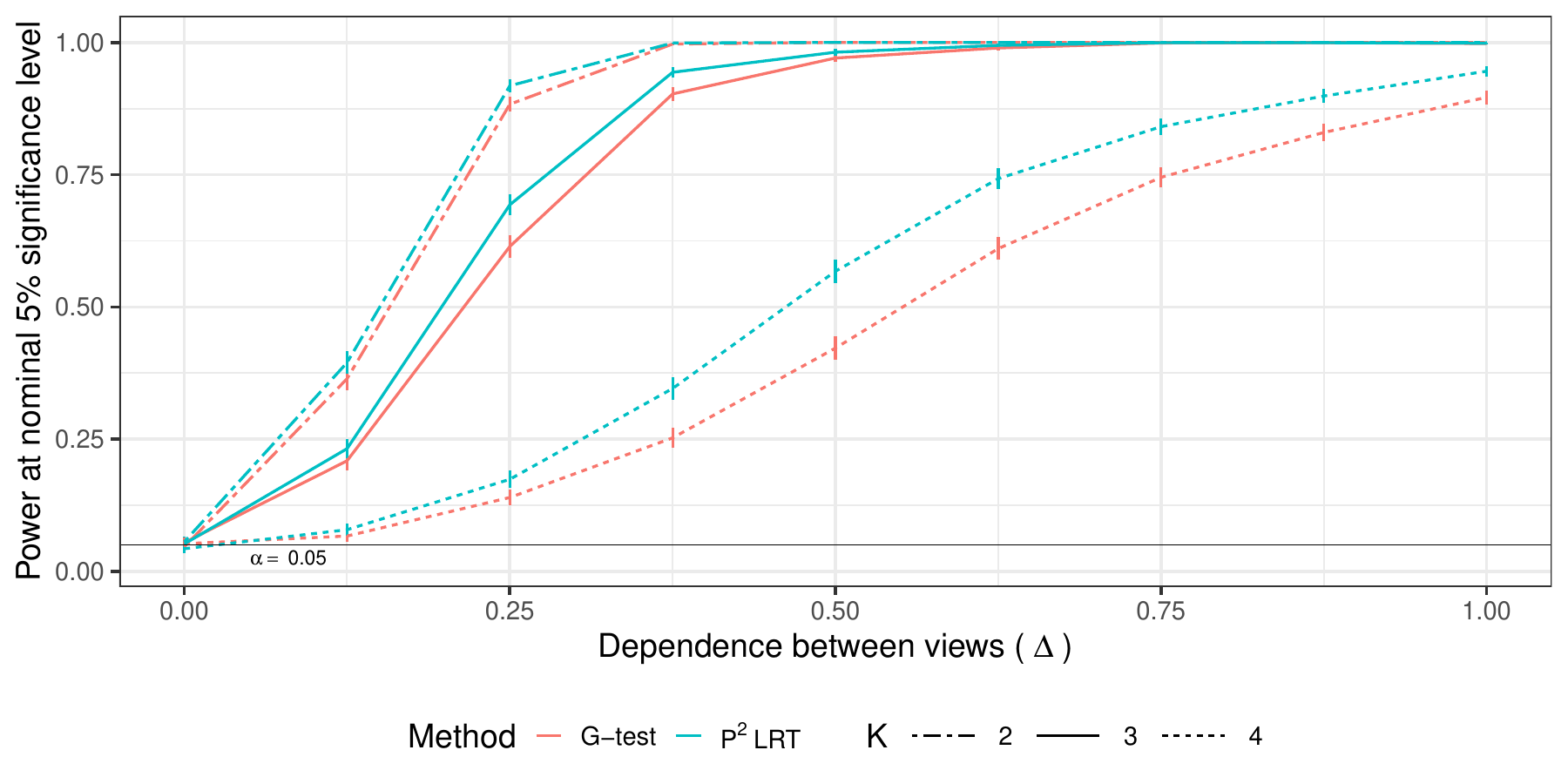}
\caption{\label{fig:s4} Power of the $P^2$LRT and the $G$-test with both views drawn from a SBM, varying the dependence between views ($\Delta$), and the number of communities in each view $(K$). The expected network density $(s)$ is fixed at $0.05$, and the community strength ($r)$ is fixed at $1.5$. Details are in Appendix \ref{sec:supp-sim2}.}
\end{figure}

We generate data from model \eqref{eq:multi-sbm-latent} --\eqref{eq:mv-subgroup-z}, with $n = 250$, $K^{(1)} = K^{(2)} = K$, $\pi^{(1)} = \pi^{(2)} = 1_K/K$, $C$ given by \eqref{eq:Csim}. We let $\thet[1] = \thet[2] = \theta$ for $\theta$ defined in \eqref{eq:thetsim}, with expected edge density $s = 0.05$, and $r = 1.5$, so that two nodes in the same community are three times more likely to be connected than two nodes in different communities. We simulate 2000 data sets for a range of values of $\Delta$ and $K$, and evaluate the power of the four tests described above. Results are shown in Figure \ref{fig:s4}. 

In Figure \ref{fig:s4}, the power of the $P^2$LRT and the $G$-test decreases as the number of communities in each view $(K$) increases. The $P^2$LRT uniformly yields higher power than the $G$-test. 

\end{document}